\documentclass[aps,pra,twocolumn,superscriptaddress, amsmath, amssymb,longbibliography]{revtex4-2}

\usepackage{graphicx}
\usepackage{dcolumn}
\usepackage{bm}
\usepackage{color}
\usepackage{todonotes}
\usepackage{xspace}

\newcommand{\FourierInt}[1]{\mathcal{I}_{#1}(\nu, \Delta)}
\newcommand{\IZero}{I^{\textrm{NR}}_{0}}
\newcommand{\ffc}{FFC\xspace}
\newcommand{\ltm}{LTM\xspace}

\allowdisplaybreaks[2]

\begin{document}

\title{Unraveling Time- and Frequency-Resolved Nuclear Resonant Scattering Spectra}

\author{Lukas Wolff}
\email[]{lukas.wolff@mpi-hd.mpg.de}

\author{J{\"o}rg Evers}
\email[]{joerg.evers@mpi-hd.mpg.de}

\affiliation{Max-Planck-Institut für Kernphysik, Saupfercheckweg 1, 69117 Heidelberg, Germany}

\date{\today}

\begin{abstract}
Owing to their extremely narrow line-widths and exceptional coherence properties, M\"ossbauer nuclei form a promising platform for quantum optics, spectroscopy and dynamics at energies of hard x-rays. A key requirement for further progress is the development of more powerful measurement and data analysis techniques. As one approach, recent experiments have employed time- and frequency-resolved measurements, as compared to the established approaches of measuring time-resolved or frequency-resolved spectra separately. In these experiments, the frequency-dependence is implemented using a tunable single-line nuclear reference absorber.
Here, we develop spectroscopy and analysis techniques for such time- and frequency-resolved Nuclear Resonant Scattering spectra in the frequency-frequency domain. Our approach is based on a Fourier-transform of the experimentally accessible intensities along the time axis, which results in complex-valued frequency-frequency correlation (\ffc) spectra . We show that these \ffc spectra not only exhibit a particularly simple structure, disentangling the different scattering contributions, but also allow one to directly access nuclear target properties and the complex-valued nuclear resonant part of the target response. In a second part, we explore the potential of an additional phase control of the x-rays resonantly scattered off of the reference absorber for our scheme. Such control provides selective access to specific scattering pathways, allowing for their separate analysis without the need to constrain the parameter space to certain frequency or time limits. All results are illustrated with pertinent examples in Nuclear Forward Scattering and in reflection off of thin-film x-ray cavities containing thin layers of M\"ossbauer nuclei.
\end{abstract}

\maketitle

\section{Introduction}
M{\"ossbauer} nuclei are established as  a versatile tool for highly-sensitive spectroscopic studies of condensed matter systems~\cite{Mossbauer1958,Kalvius2012_book,Yoshida2021}. Owing to their extremely narrow line-widths and exceptional coherence properties~\cite{Smirnov1999,HannonTrammell1999, Kagan1999}, they also form a promising platform for quantum optics at hard x-ray energies~\cite{adams_x-ray_2013,kuznetsova_quantum_2017,Adams2019,jaeschke_quantum_2020,yoshida_quantum_2021}. In the future, state-of-the-art and upcoming x-ray light sources with unprecedented quality and intensity may open up a regime of nonlinear~\cite{https://doi.org/10.48550/arxiv.1607.04116,PhysRevResearch.4.L032007,RevModPhys.69.1085} and non-equilibrium phenomena \cite{Shenoy2008} which has only started to be explored experimentally \cite{chumakov_superradiance_2018}. However, exploring this  regime will require advanced measurement and data analysis techniques to access a broader range of observables, and to compare theoretical predictions with experimental observations. Examples include spectroscopy beyond the linear response regime~\cite{Mukamel1995}, photon-correlation measurements~\cite{PhysRev.130.2529}, or methods to study the time-resolved sample dynamics after external stimuli, potentially on a per-shot basis~\cite{Shenoy2012}.

\begin{figure*}[t]
\includegraphics[width=0.8 \textwidth]{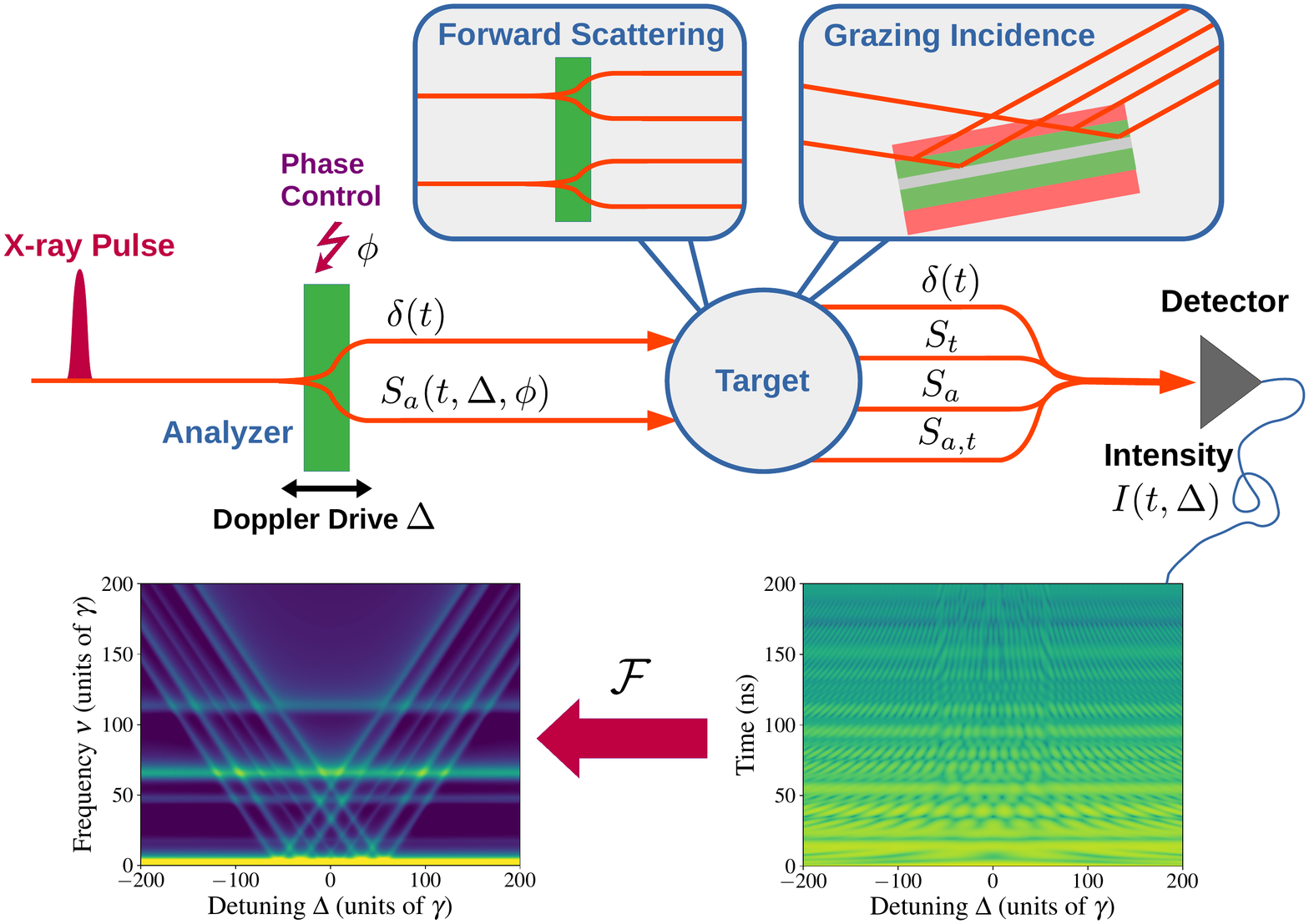}
\caption{\label{fig.Setup} Schematic of a setup to record time- and frequency-resolved Nuclear Resonant Scattering spectra in forward and grazing incidence geometry. The goal is to characterize the resonant target response. An additional single-line analyzer mounted on a Doppler drive is used to introduce a variable frequency dependence.  Resonant and non-resonant scattering in analyzer and target lead to four interfering scattering pathways contributing to the detection signal. An example for the experimentally accessible scattered intensity as function of analyzer frequency and arrival time of the scattered photons is shown in the lower right panel. The corresponding frequency-frequency correlation (\ffc) spectrum studied in this work is obtained via a Fourier transform along the time axis.  An example \ffc spectrum for the data in the lower-right panel is shown in the lower-left panel. It exhibits clear horizontal and diagonal linear structures which can be interpreted in terms of spectral correlations within the combined analyzer-target-system. In the second part of this work, an additional phase control between resonant and non-resonant response of the analyzer is considered to disentangle the different scattering pathways. }
\end{figure*}

As a first step towards this goal, recent experiments employed a time- and frequency-resolved measurement of the scattered x-ray intensity~\cite{Heeg2017,Heeg2021,Shvydko2022,Heeg2022}, as compared to the established approaches of measuring time-resolved or frequency-resolved spectra separately~\cite{Kalvius2012_book}, the latter of which can be obtained, for instance, by probing the nuclear absorption using pure nuclear Bragg reflections to produce highly monochromatic x-ray light on the scale of the nuclear line-width \cite{Smirnov1997}. Alternatively, the frequency-selectivity can be achieved using a heterodyne approach by adding an additional single-line reference absorber, in the following referred to as analyzer, on a velocity drive  up- or downstream of the target under investigation~\cite{PhysRevB.54.16003,Neyens1998}, see Fig.~\ref{fig.Setup}. This method is also used to perform time- and frequency-resolved data acquisition, which provides a number of significant advantages over other detection schemes using single-line reference absorbers, even though it does not require changes to the experimental setup apart from the electronics.
On the one hand, it allows one to apply several established data analysis methods using a single experimental data set only. For example, late-time integration methods established as a standard analysis approach for x-ray cavities probed in reflection~\cite{PhysRevB.54.16003,Deak2006,Roehlsberger2010} or related stroboscopic methods~\cite{PhysRevB.65.180404,PhysRevB.67.104423, Roehlsberger2012} can be employed by integrating the two-dimensional data set along part of the time axis. Importantly, the time- and frequency-resolved spectra allow one to improve the recovery of the target spectra by optimizing the integration range throughout the data analysis~\cite{Coussement2000,Herkommer2020}. Similarly, off-resonant methods can be employed in spectral regions with large detuning between analyzer and target, which may even provide access to the complex-valued target response~\cite{Callens2005,PhysRevB.65.180404,PhysRevB.67.104423,Goerttler2019}. These methods typically exploit the interference between particular scattering pathways, which can be studied in the time~\cite{Callens2005}- or frequency domain~\cite{Goerttler2019}. However, these established methods share the drawback that they only make use of select regions of the recorded two-dimensional spectra.
On the other hand, the two-dimensional data set allows for a much more stringent comparison between theory and experiment. One reason for this are the rich interference structures, which are lost in the usual one-dimensional data by the integration over the other axis (see bottom-right panel of Fig.~\ref{fig.Setup} for an example). Interestingly, these structures encode intensity and phase of the target response. In Refs.~\onlinecite{Heeg2017,Heeg2021}, the complex-valued target response was determined by fitting theory models to the entire two-dimensional spectrum, thereby using all recorded data rather than only parts of it. However, this approach is computationally demanding as compared to other methods, and requires model fits in order to extract the desired target properties.

This raises the question, if the two-dimensional spectra can also be analyzed in different ways, which ideally provide access to the desired target properties in a more transparent way, without requiring global fits to the entire spectra, but still allow one to exploit the time- and energy correlations in the spectra, and to make use of large parts or even the entire experimental data set in the analysis.

Motivated by this, here, we develop spectroscopy and analysis techniques which are based on the Fourier transform of experimentally accessible time- and frequency-resolved intensities along the time axis.  The resulting complex-valued frequency-frequency correlation (\ffc) spectra exhibit particularly simple signatures comprising horizontal and diagonal structures, which can be associated to different contributing scattering processes. These signatures (see the bottom-left panel of Fig.~\ref{fig.Setup} for an example) facilitate a selective analysis of the different scattering contributions. We in particular focus on two analysis approaches. First, we discuss linear fits to the diagonal structures in the frequency-frequency correlation spectra, which allow one to extract nuclear resonance energies, as well as spectral line features such as collective energy shifts and superradiant line broadenings.  Second, we show that horizontal or vertical sections through the diagonal structure  provide access to amplitude and phase of the nuclear resonant part of the target response, cross-correlated with the analyzer response. This retrieval of the response without contributions from the off-resonant scattering in the target is of particular interest for x-ray cavity targets containing M\"ossbauer nuclei, since their spectra typically are strongly affected by the interference of electronic and nuclear response, in dependence on the x-ray incidence angle.
We further show that an additional control of the relative phase between the electronic and nuclear response of the analyzer allows one to disentangle different scattering pathways, thereby facilitating their selective analysis without imposing additional constraints such as a large analyzer-target detuning. All approaches are illustrated using examples of practical relevance.

The manuscript is structured as follows: The next Section briefly describes a generic experimental setup used to record time- and frequency-resolved spectra including phase- and frequency-control of the nuclear reference absorber. Further, we derive expressions for the frequency-frequency correlation spectra in linear response theory, and discuss them in particular limiting cases.
Section~\ref{sec.Diagonals} presents the two analysis approaches for the diagonal structures, including numerical examples in forward scattering and cavity reflection. Section~\ref{sec.PhaseCycle} introduces the phase control of the analyzer as an additional control parameter, and discusses its implications for the analysis of the diagonal structure. Finally, Sec.~\ref{sec.summary} summarizes the results.

\section{Linear-response formalism and spectral correlations in nuclear resonant scattering\label{sec:theory}}

For our analysis, we consider the setup shown schematically in Fig.~\ref{fig.Setup}. A temporally short and spectrally broad x-ray pulse delivered by an accelerator-based x-ray source is used to probe a target containing M\"ossbauer nuclei. Both, forward scattering geometries and reflection from x-ray cavities will be considered. Additional frequency information is gained by introducing a single-line reference absorber, which can be tuned in frequency by $\Delta$ via a Doppler drive. Throughout this paper, frequencies are given in units of the target single-nucleus line-width $\gamma$.
Each target features a near-instantaneous electronic response $\propto \delta(t)$, and a delayed nuclear response, which we denote as $S_i(t)$, with $i\in\{a,t\}$ for analyzer and target~\cite{lynch_time_1960,kagan_excitation_1979,Smirnov1999,Shvydko1998, RoehlsbergerBook,HannonTrammell1999}. The two-stage setup thus gives rise to four different scattering channels~\cite{PhysRevA.71.023804,Callens2005}, as illustrated in Fig.~\ref{fig.Setup}.
In the experiment, the time- and frequency-resolved intensity of the scattered light is measured, which gives rise to two-dimensional spectra as illustrated in the bottom right panel of Fig.~\ref{fig.Setup}~\cite{Heeg2017,Heeg2021}. A Fourier transform along the time axis then leads to the frequency-frequency correlation spectra considered here. We note that Ref.~\onlinecite{Goerttler2019} employed a similar Fourier transform in order to select a particular frequency region for a subsequent analysis in the time domain. The bottom left part of Fig.~\ref{fig.Setup} shows the real value as an example, clearly exhibiting the horizontal and diagonal structures.
In Sec.~\ref{sec.PhaseCycle}, we will further consider the possibility of controlling the relative phase $\phi$ between the electronic and nuclear response of the analyzer.

\subsection{Time- and frequency resolved spectra in the linear response formalism}

A theoretical description of time- and frequency-resolved Nuclear Resonant Scattering spectra can be given employing the linear response formalism (see e.g. \cite{Smirnov1999,HannonTrammell1999}) which is justified by the narrow line-width characteristic of M{\"o}ssbauer transitions that typically leads to low excitation in state-of-the-art experiments, even at high-brilliance third-generation synchrotron sources. Neglecting polarization effects~\cite{Siddons1999}, each target $i$ in the beam path can be described by a scalar transfer function $\hat{T}_i(\omega)$ in the frequency domain or (impulse) response function $T_i(t)$ in the time domain. Here and in the following, the ``hat'' denotes quantities in the frequency domain. Then, the outgoing field is given by
\begin{align}
\hat E_{out}(\omega) &= \hat{T}_i(\omega) \hat E_{in}(\omega) \,,\\
 E_{out}(t) &= (T_i \ast E_{in})(t) \label{eq.eout}
\end{align}
in the frequency- or time domain, respectively. Next to the convolution $\ast$ in Eq.~(\ref{eq.eout}), we also define the cross-correlation $\star$,
\begin{align}
(f \ast g)(x) & = \int^{\infty}_{-\infty}dyf(x-y)g(y) \label{eq.Convolution}\,,\\
(f \star g)(x) & = \int^{\infty}_{-\infty} dy f^*(y-x)g(y) \label{eq.Correlation}\,,
\end{align}
for two complex-valued functions $f,g$ of frequency or time variables $x,y$. The convolution can be interpreted as applying a filter $f$ to $g$, or a propagation of an input $g$ at point $y$ to the final output $f \ast g$ at point $x$ by virtue of the response function $f$. The cross-correlation $f \star g$ on the other hand can be thought of as scanning the functions $f$ and $g$ for similarities by introducing relative shifts $x$ between them. Both quantities and their interpretations will turn out to be instrumental for understanding the diagonal and horizontal structures appearing in Fourier-transformed time- and frequency-resolved spectra, the real part of which is shown in the lower left plot of Fig.~\ref{fig.Setup} as an example.

The responses of nuclear targets in forward scattering and grazing incidence geometry comprise two fundamentally different scattering processes: Prompt scattering nonresonant with the nuclear transition and coherent resonant scattering delayed by the slow decay of the nuclear transition. On the time scale of the nuclear decay, the prompt radiation can be described by a Dirac-$\delta(t)$-like pulse and thus the outgoing field in Nuclear Resonant Scattering is of the form 
\begin{align}
E_{out}(t) = \alpha[ \delta(t) + S_i(t)] \ast E_{in}(t)\,.\label{eq:gen-time}
\end{align}
Here, the prefactor $\alpha$ accounts for attenuation and dispersion imposed by the surrounding nonresonant material and $S_i(t)$ denotes the nuclear resonant part of the target's response. A two-target setup formed by a  reference absorber foil  $T_a$ (analyzer) and an unknown target $T_t$ in forward scattering geometry as depicted in Fig.~\ref{fig.Setup} can be described by the total response function $T(t) = (T_t \ast T_a)(t)$. The response of the reference absorber with tunable transition frequency $\omega_a + \Delta$ and relative phase $\phi$ between the prompt and scattered part can be written as
\begin{align}
 T_a(t,\Delta, \phi) = \alpha_a \left[\delta(t) + e^{-i\Delta t}e^{i\phi}S_a(t)\right]\,.
\end{align}
Note that, typically, we will consider thin reference absorbers whose spectral features are more narrow than those of the target absorber. However, the subsequent analysis does not employ approximations of the reference absorber's response function based on this thin-analyzer limit, and our numerical results below will exhibit effects beyond this limit. In the following, for notational brevity, we will absorb the detuning and phase dependence into the nuclear scattering response as
\begin{align}
 S_a(t,\Delta, \phi) = e^{-i\Delta t}e^{i\phi}S_a(t)\,,
\end{align}
and suppress the dependence on $\phi$ throughout this Section as phase control will become of relevance only later in Sec.\ref{sec.PhaseCycle}.

With these considerations, the experimentally accessible time- and frequency-dependent intensity at the detector can be expressed in terms of response functions as
\begin{align}
 I(t, \Delta) &= |(T \ast E_{in})(t)|^2 \nonumber \\[2ex]
 &= (T \ast E_{in})^*(t)\: (T \ast E_{in})(t)\,, \label{eq.tf-int}
\end{align}
where the $\Delta$-dependence arises via $T_a$ in the response function $T$. Such two-dimensional time- and frequency-resolved spectra allow for a much more stringent comparison of experimental data to theory predictions than the corresponding one-dimensional time-spectra or energy-spectra alone, and have been measured in recent experiments~\cite{Heeg2013SGC,Heeg2017,Heeg2021}

\subsection{Frequency-frequency correlation spectrum \label{sec.Correla}}

In order to discuss spectral correlations, we define the frequency-frequency correlation (\ffc) spectrum as the Fourier-transform of the experimentally accessible intensity Eq.~(\ref{eq.tf-int}) along the time axis, 
\begin{align}
\FourierInt{} &=\int^{\infty}_{-\infty}dt \: e^{i\nu t}\: I(t, \Delta) \label{eq.InfTimeIntensity} \\[2ex]
& = E^2_0 \int^{\infty}_{-\infty}dt \: e^{i\nu t} \: T^*(t)\:T(t) \nonumber \\[2ex]
& = \frac{E^2_0}{2\pi} (\hat{T} \star \hat{T})(\nu)\,, \label{eq.SpecAuto}
\end{align}
where the initial pulse was written as $E_{in}(t) = E_0 \delta(t)$ as it is typically orders of magnitude shorter than the nuclear evolution time scales. Note that $I(t,\Delta)$ vanishes for times $t<0$ since the excitation occurs at $t=0$. However, the symmetric integration will allow us to derive simple analytical expressions for the case with detection time gating in Sec.~ III D. Interestingly, the \ffc spectrum Eq.~(\ref{eq.InfTimeIntensity}) can be expressed as the (spectral) auto-correlation of the total response function $\hat T$.  The result Eq.~(\ref{eq.SpecAuto}) can thus be regarded as a frequency-domain instance of the Wiener-Khinchin theorem which relates the Fourier transform of the power spectral density $\lvert T(t)\rvert^2$ to the autocorrelation $(\hat{T} \star \hat{T})(\nu)$ (see, e.g., \cite{EngelbergBook}). As we will see in the following, the Fourier transformation in Eq.~(\ref{eq.InfTimeIntensity}) translates temporal interference effects into spectral correlations from which spectral features and the phase information of the nuclear target can be extracted.
However, it is important to note that the \ffc spectrum itself is not an intensity, as it is complex-valued for general Fourier frequencies $\nu$. The reason is that it is derived via the Fourier transform from the experimentally accessible real-valued intensity. Regarding previous detection schemes using single-line reference absorbers as mentioned in the introduction, we note that the \ffc spectrum reduces to the real-valued time-integrated spectrum for $\nu = 0$ (cf. \cite{PhysRevB.54.16003}). By only integrating over certain parts of the time-axis, late-time and stroboscopic spectra (see, e.g. \cite{PhysRevB.67.104423}) can be recovered from the same $\nu = 0$ contribution. In this sense, the \ffc spectrum can be regarded as a generalization of these methods to arbitrary Fourier frequencies $\nu$ using the same data set. The case where the signal at early times is discarded, is discussed in more detail in Sec.~\ref{subsec.Gating} and appendix~\ref{app:late}. In the following, we focus on the ideal case in which all times are available for data analysis in order to derive analytical expressions for the most relevant features of these spectra.
\begin{figure*}[t]
\includegraphics[width = \textwidth]{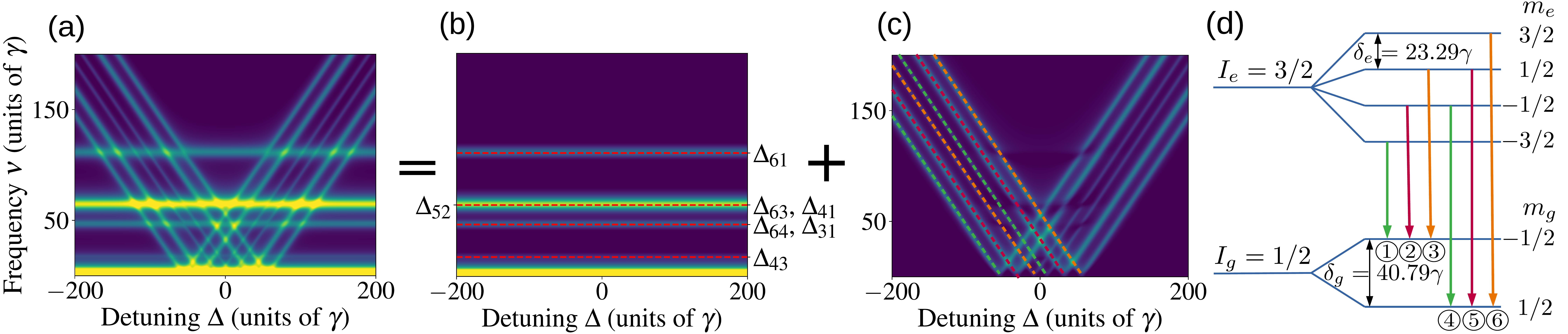}

\caption{Decomposition of \ffc spectra into single-target and two-target contributions. (a) shows the real part of the \ffc spectrum $\FourierInt{}$ after removal of the off-resonant background. It can be separated into two parts shown in (b) and (c). The first part in (b) is the sum of the individual responses of target and analyzer, corrected for the non-resonant absorption. The horizontal structures are due to quantum beats in the target.  As guide to the eye, the relevant mutual detunings  between the target transitions are indicated by dashed red lines. (c) is the difference between the spectra in (a) and (b). This spectrum is dominated by diagonal structures, which originate from the interference of the resonant scattering off of the different target resonances with the resonant analyzer response.
These results are obtained for a 2$\mu$m thick enriched $\alpha-{}^{57}$Fe target with hyperfine field $B=33.3$~T and a stainless-steel single-line analyzer as described in the main text. The level structure and the relevant transitions in the target are shown in (d). The orange, red and green transitions describe the scattering of left-circularly, linearly and right-circularly polarized light, respectively. \label{fig.Unravel} }
\end{figure*}

A numerical example for the \ffc spectrum in Eq.~(\ref{eq.InfTimeIntensity}) is given in Fig.~\ref{fig.Setup}. The bottom-right panel shows the experimentally accessible time- and frequency-resolved intensity $I(t, \Delta)$ in Eq.~(\ref{eq.tf-int}). The bottom-left panel shows the real part of the \ffc spectrum in Eq.~(\ref{eq.InfTimeIntensity}), which is dominated by a set of horizontal and diagonal spectral features in the $\Delta\!-\!\nu$-plane. Note, that throughout this paper, we only plot and analyze the positive $\nu$ branch of the \ffc spectrum since the inversion of the Fourier frequency $\nu \rightarrow -\nu$ leads to complex conjugation of the \ffc signal. The main part of our \ffc analysis, however, will be carried out on its real part or absolute value and thus the negative $\nu$ branch contains only redundant information.
 
For an interpretation of these diagonal features, we rewrite the spectral auto-correlation function Eq.~(\ref{eq.SpecAuto}) using the Fourier-domain response functions 
\begin{align}
 \hat{T}(\omega) &= \hat{T}_t(\omega)\hat{T}_a(\omega)\,,\\
 \hat{T}_a(\omega, \Delta, \phi) &= \alpha_a\left[1+\hat{S}_a(\omega,\Delta, \phi)\right] \label{eq.FourierResp}
\end{align}
as
\begin{align}
\FourierInt{} = & \frac{E^2_0 \lvert \alpha_a \rvert^2}{2\pi} \left[(\hat{T}_t \star \hat{T}_t)(\nu)+ (\hat{T}_t \star \hat{T}_t \hat{S}_a)(\nu) \right. \label{eq.AnaForm}  \\
& \left. +(\hat{T}_t\hat{S}_a \star \hat{T}_t)(\nu)+ (\hat{T}_t\hat{S}_a \star \hat{T}_t \hat{S}_a)(\nu)\right] \; . \nonumber
\end{align}
Eq.(\ref{eq.FourierResp}) shows the clear separation of nonresonant (electronic) scattering, which is approximately constant on the scale of the nuclear resonances, and the frequency-dependent nuclear resonant scattering in form of the nuclear resonant response $S_a(\omega,\Delta, \phi)$ in the frequency-domain. This separation is crucial for the evaluation of the \ffc spectrum in the form of Eq.(\ref{eq.AnaForm}), the single parts of which can be interpreted as follows: the first term describes spectral correlations between different transitions in the target, as will be further discussed below.  The other three terms contain the contribution $\hat{T}_t\hat{S}_a = \alpha_t [\hat{S}_a +  \hat{S}_a \hat{S}_t ]$ combining target and analyzer. Its first addend $\sim\hat{S}_a$ arises from the aforementioned nonresonant (electronic) scattering in the target followed by resonant (nuclear) scattering processes in the analyzer foil. It forms the basis of the heterodyne-type detection schemes which determine the target response using interference between radiation emitted from the target and analyzer, respectively (cf. \cite{Callens2005, Coussement2000}). It will turn out to be the main origin of the diagonal structure found in \ffc spectra. The second addend $\sim \hat{S}_a \hat{S}_t = \hat{S}_{a,t}$, known as radiative coupling \cite{Smirnov1999, HannonTrammell1999,PhysRevA.65.023804,PhysRevA.71.023804}, describes processes with resonant scattering in both analyzer and target.
In the following, we will exploit that these two scattering contributions are naturally separated owing to their different scattering amplitudes as a function of detuning between analyzer and target and will focus on the first contribution, as it dominates the \ffc spectra in the large analyzer-target detuning limit.

\subsection{Large analyzer-target detuning limit\label{sec:large-det}}
The resonantly scattered part of the analyzer response $\hat{S}_a(\omega)$ is nonzero in the vicinity of the resonance frequency $\omega = \omega_a +\Delta$ only. For the same reason, the full target response becomes spectrally flat far-off nuclear resonance, i.e., $\hat{T}_t(\omega) \approx \alpha_t$. As a result, we can approximate
\begin{align}
 \hat{T}_t(\omega) \hat{S}_a(\omega) \approx \alpha_t \hat{S}_a(\omega)
\end{align}
in the limit of large detunings $\Delta +\omega_a-\omega_t$ between the analyzer and target transitions with frequencies $\omega_t$. In this approximation, the spectral auto-correlation function can be written as
\begin{align}
\FourierInt{\mathrm{off}}
 &\approx  \frac{E^2_0 \lvert \alpha_a\rvert^2}{2\pi} \left[
 (\hat{T}_t \star \hat{T}_t)(\nu) 
+ \lvert \alpha_t\rvert^2 ( \hat{S}_a \star \hat{S}_a )(\nu)
 \right. \nonumber  \\
& \qquad\left.
 + \alpha_t  (\hat{T}_t \star \hat{S}_a)(\nu)%
+ \alpha^*_t(\hat{S}_a \star \hat{T}_t) (\nu) 
\right]\,. \label{eq.IntOff}
\end{align}
The first two terms correspond to the full target response $\mathcal{F}[\lvert T_t(t)\rvert^2]$ and the resonantly scattered part of the analyzer response $\mathcal{F}[\lvert S_a (t)\rvert ^2]$, where $\mathcal{F}$ denotes the Fourier transform. These contributions can be determined from separate measurements of the time-dependent scattered intensity of the target and the analyzer alone, i.e., in the absence of the respective other target. Their complex-valued contributions to Eq.~(\ref{eq.IntOff}) then follow from a Fourier transformation analogous to Eq.~(\ref{eq.InfTimeIntensity}). 

This separation of the single-stage contributions $\hat{T}_t \star \hat{T}_t$ and $\hat{S}_{a} \star \hat{S}_a$  from the \ffc spectrum is illustrated in Fig.~\ref{fig.Unravel}. The full \ffc spectrum in panel (a) decomposes into the single-stage response parts shown in (b) and the residual two-target contributions in (c). Clearly, the single-stage responses correspond to the horizontal lines in (a) since they do not depend on $\Delta$, whereas the two-target parts give rise to the diagonal lines. Note that, if the single-stage contributions are removed by subtracting the time-resolved analyzer and target intensities, the count rates of the three measurements of the time- and frequency-resolved spectrum, the time-resolved target spectrum and the time-resolved analyzer spectrum, have to be adjusted to each other, e.g. by adapting the count rates such that after subtraction the horizontal lines have been removed. Alternatively, the horizontal lines can be removed directly from the \ffc spectrum by determining their vertical position and shape far away from the diagonal structure. As these contribution to the \ffc spectrum are detuning-independent, a subtraction of the result directly clears the horizontal lines from the entire spectrum without the need of additional measurements. We pursue the first approach in the numerical analysis in Sec.~\ref{sec.Diagonals}, while the second one is applied to the horizontal lines appearing in the sum spectrum $\mathcal{I}_{inv}$ in Sec.~\ref{sec-phasediag}.

After subtracting the single-stage contributions, we remove the constant off-resonant background contribution to $\hat{T}_t$, which can be written in the same form as Eq.~(\ref{eq.FourierResp}), i.e. the part proportional to $ \lvert \alpha_t \rvert^2$ in the second line of Eq.~(\ref{eq.IntOff}), to study the nuclear resonant part $\hat{S}_t$ of the response alone. This can be done, e.g. by rejecting the first few nanoseconds after pulse arrival at the detector. Then, finally, the spectral auto-correlation function in the large analyzer-target detuning limit Eq.~(\ref{eq.IntOff}) can be written as
\begin{align}
\FourierInt{Dia} =& \frac{\IZero}{2\pi}\left[(\hat{S}_a \star \hat{S}_t)(\nu)+ (\hat{S}^*_a \star \hat{S}^*_{t})(-\nu)\right]\,, \label{eq.OffResult}
\end{align}
where we used $(f \star g)(\nu) = (g^* \star f^*)(-\nu)$. Further, we defined   $\IZero = E^2_0 \lvert \alpha_a \alpha_t\rvert^2$ which describes the intensity at the detector after the two-target setup in the absence of nuclear scattering. 

It is important to note that throughout the derivation of Eq.~(\ref{eq.OffResult}), no assumptions were made on the analyzer and target response functions, except for the general features that the scattering contributions to the response functions cover only a limited spectral region, and that the full response function is spectrally flat outside this nuclear resonance region. Therefore, the expression Eq.~(\ref{eq.OffResult}) holds for general nuclear targets.

The interpretation of Eq.~(\ref{eq.OffResult}) will be discussed in more detail in Sec.~\ref{sec:int-ffc} which also reveals its relation to the diagonal structures in the \ffc spectrum and how it can be employed to extract the complex-valued nuclear resonant part of the target response from experimentally accessible \ffc spectra.

\subsection{\label{sec:offres}Relation to off-resonant methods in the time domain}

The derivation of Eq.~(\ref{eq.OffResult}) relied on the large analyzer-target detuning, in order to isolate a single scattering channel of interest. In the literature, several methods have been reported which employ a similar parameter regime in order to study time- and frequency-resolved Nuclear Resonant Scattering spectra, specifically for the recovery of the complex-valued target response function \cite{Callens2005,Goerttler2019}. These latter methods operate in the time-domain, and we can connect our energy-domain result Eq.~(\ref{eq.OffResult}) to them via the so-called cross-correlation theorem \cite{WeissteinWolfram} as
\begin{align}
\FourierInt{RM} =& \frac{\IZero}{2\pi}\left[(\hat{S}_t \star \hat{S}_a)(\nu)+(\hat{S} _a \star \hat{S}_{t})(\nu)\right] \nonumber \\
=& \frac{\IZero}{2\pi}\Big\{\Big(\mathcal{F}[S^*_t (t)] \ast \mathcal{F}[S_a(t)]\Big)(\nu) \nonumber \\
&+\Big(\mathcal{F}[S^*_a(t)] \ast \mathcal{F}[S_t(t)] \Big)(\nu)\Big\} \nonumber \\
=& 2\IZero \int^{\infty}_{-\infty}dt e^{i\nu t}\textrm{Re}\left[S^*_t(t)S_a(t)\right]\,. \label{eq.CrossCorr}
\end{align}
We see that the spectral cross-correlation between target and analyzer response are mapped to the interference between the corresponding time-domain responses and thus, studying the diagonal structure in the \ffc spectrum is analogous to studying the hyperbolic interference pattern appearing in the time- and frequency-resolved intensities.

Our analysis below exploits specific features of the energy-domain representation, and extends the previous results in several ways. Most notably, we will discuss the spectroscopy of thin-film cavities probed in grazing incidence, and show that the off-resonant approaches allow one to access the nuclear target response alone, without the usual interference with the electronic scattering on the cavity structure.

Further, it is important to note that both the energy- and time-domain approaches discussed so far are restricted to off-resonant spectral regions, in which the radiative coupling contributions are negligible. Hence, data recorded near resonance between analyzer and target cannot be employed for the analysis and, as a consequence, these spectroscopy methods do not allow one to access the interesting regime in which radiative couplings substantially modify the target dynamics~\cite{PhysRevA.65.023804,PhysRevA.71.023804,Heeg2021,Shvydko2022,Heeg2022}. To overcome these limitations, below in Sec.~\ref{sec.PhaseCycle}, we will discuss a method to suppress different scattering channels to become more independent of the large analyzer-target detuning limit or related approximations, and will demonstrate its advantages by comparison to the off-resonant method.

A brief comparison of the off-resonant methods to the so-called late-time integration spectroscopy methods is given in Appendix~\ref{app:late}.

\subsection{Interpretation of the frequency-frequency correlation spectrum \label{sec:int-ffc}}
After having derived the spectral auto-correlation function Eq.~(\ref{eq.SpecAuto}) in the large analyzer-detuning limit Eq.~(\ref{eq.OffResult}), and hence a representation of the \ffc spectrum in terms of response functions in this limit, we now turn to its interpretation, which will form the basis for further analysis below.
\subsubsection{Thin-analyzer limit}
To this end, we for the moment assume an idealized case of a single-line analyzer with negligible line-width and absorption $\alpha_a$ and resonance frequency $\omega_a+\Delta$ given by the scattering response function
\begin{align}
 \hat{S}_a(\omega, \Delta) = \delta(\omega-\Delta-\omega_a).
\end{align}
Then, Eq.~(\ref{eq.OffResult}) simplifies to 
\begin{align}
\FourierInt{\textrm{id}} = \frac{\IZero}{2\pi}&\left[ \hat{S}_t(\omega_a+\Delta+\nu) \right. \nonumber \\
&\left. +\hat{S}^*_t(\omega_a+\Delta-\nu)\right]\,. \label{eq.trideal}
\end{align}
Thus, in this idealized (``id'') analyzer case, the \ffc spectrum provides direct access to the desired nuclear resonant target response $\hat{S}_t$. Interestingly, the two contributions of  $\hat{S}_t$ in Eq.~(\ref{eq.trideal}) are centered around $\omega_a+\Delta \pm \nu$. This can be understood by noting that the analyzer response at $\omega_a+\Delta$ is shifted by the Fourier frequency $\pm \nu$ in the cross correlation. We can thus identify the two contributions as the origin of the two branches of the diagonal structures with positive or negative slopes in the \ffc spectrum. 

Below, we will also consider analyzers with finite thickness, in which case Eq.~(\ref{eq.trideal}) generalizes to Eq.~(\ref{eq.OffResult}), i.e., the target response is additionally cross-correlated with the analyzer response function.

As a final remark, we note that, while the diagonal structure is directly related to the nuclear resonant target response $\hat{S}_t$, the \ffc spectrum in principle also allows to study the full target response $\hat{T}_t$ if the nonresonant background is not removed in the derivation of Eq.(\ref{eq.OffResult}), at the cost of retaining a large nonresonant background in the \ffc spectrum. This approach, however, is out of the scope of this publication.

\subsubsection{\label{sec:extract}Extracting information on the target response from the diagonal structures in the \ffc spectrum}
As a result of Eq.~(\ref{eq.trideal}), the target can be characterized via the diagonal structures in the \ffc spectrum in two different ways. First, the diagonals are governed by the relation 
\begin{align}
 \nu(\Delta) = \pm(\Delta -\omega_t+\omega_a)\,, \label{eq.diag}
\end{align}
which defines lines in the $\nu$-$\Delta$-plane with slope $\pm 1$ and offset $\pm(\omega_a-\omega_t)$. Thus, upon extrapolating the diagonals in the \ffc spectrum, they both cross the $\nu = 0$ coordinate axis at $\Delta =  \omega_t - \omega_a$, and thereby provide access to the target transition frequency. Note that this argument generalizes to multiple target transitions as discussed below.

Second, one may analyze the spectral shape of the diagonals in the \ffc spectrum as function of $\Delta$ or $\nu$. From Eq.~(\ref{eq.trideal}) it follows that in both cases, this shape corresponds to the desired shape of the nuclear resonant target response. Note that the \ffc spectrum is complex-valued, and thereby provides access to the amplitude and the phase of the resonant response $\hat{S}_t$. This ``phase problem'' of extracting the phase of a target response is an ubiquitous problem in spectroscopy, imaging and diffraction experiments \cite{Elements,Taylor:ba5050,Sturhahn2001,Callens2005}.

Both approaches will be discussed in Sec.~\ref{sec.Diagonals}. There, the more realistic situation with a finite analyzer width in Eq.~(\ref{eq.OffResult}) will be considered. Then, the target spectrum is broadened and modified in amplitude through the correlation with the analyzer spectrum. Therefore, in practice, analyzers with a spectrum narrow as compared to the desired target spectrum should be employed. Further, different horizontal and vertical cuts through the diagonals can be combined in order to reduce uncertainties in the recovery of the target response.  

\subsubsection{Horizontal structures in the \ffc spectrum \label{sec.Horizontal}}

The horizontal lines appearing in the \ffc spectrum (c.f. Fig.~\ref{fig.Unravel}) can be attributed to a beating of light scattered on  target transitions with different resonance frequencies. The different scattering channels interfere due to the coherent nature of the scattering, giving rise to the so-called quantum beats~\cite{Smirnov1999,HannonTrammell1999,RoehlsbergerBook}. As in the case of the diagonal structures in the \ffc spectrum, the Fourier-transform of the time- and frequency resolved intensity of this beating pattern will peak whenever the Fourier frequency $\nu$ equals one of the detunings between a pair of interfering hyperfine transitions (Fig.\ref{fig.Unravel} (b)). Note, however, that not all detunings between hyperfine transitions, which are indicated in Fig.\ref{fig.Unravel} (c), appear as horizontal lines in the spectrum. This is due to the fact that in our calculation the sample is irradiated by an unpolarized beam and the hyperfine field is oriented perpendicularly to the beam direction. Hence, photons scattered off the $\Delta m = 0$ and $\Delta m = \pm 1$ are orthogonally polarized and do not interfere with each other (cf. \cite{Siddons1999}) and thus the corresponding peaks/lines do not appear in Fig.\ref{fig.Unravel}. 

\section{Analysis of the diagonal structure\label{sec.Diagonals}}

In the previous Sec.~\ref{sec:extract}, we have shown that the diagonal structure in the \ffc spectrum is formed by correlations of analyzer and target responses centered around positions $\nu(\Delta) = \pm(\Delta -\omega_t+\omega_a)$. This allows for the determination of spectral target parameters by extrapolating the diagonals towards the crossing point with the $\nu$- or $\Delta$-axis. Furthermore, horizontal or vertical cuts through the diagonals provide access to the complex-valued nuclear resonant part of the target response function $\hat{S}_t$ (cross-correlated with the analyzer response function). In this Section, we demonstrate the practical feasibility of both of these  approaches, by evaluating representative model data of nuclear forward scattering targets and thin-film cavities.

\subsection{Computational details on the analysis\label{sec.comp-details}}

The model data evaluated in this Section are calculated using the software package \textsc{pynuss} \cite{Pynuss}, which features methods to evaluate  Nuclear Forward Scattering and cavity reflection spectra, similar to  the software package \textsc{conuss} \cite{Sturhahn2000}, but is written in the language Python~\cite{10.5555/1593511} and features substantial extensions for the analysis of quantum optical applications in Nuclear Resonant Scattering. In \textsc{pynuss}, forward scattering and cavity reflection spectra are calculated using the transfer matrix and layer formalism, respectively~\cite{RoehlsbergerBook,Sturhahn2000}. Our model system is the M{\"o}ssbauer isotope $^{57}\textrm{Fe}$ with transition energy $\hbar \omega_t = 14.4$keV and single-nucleus decay rate $\hbar \gamma = 4.7$neV. As analyzer, an enriched stainless steel sample ($^{57}\textrm{Fe}_{\, 55}\textrm{Cr}_{25}\textrm{Ni}_{20}$) of thickness $1\mu$m with the same transition frequency $\omega_a = \omega_t$ (the isomeric shift is neglected for simplicity, without loss of generality) and single-nucleus line-width is used for the calculation. Different targets will be considered in the analysis, as specified below.

Following Sec.~\ref{sec:theory}, the diagonal analysis is then performed on the complex-valued quantity
\begin{align}
\mathcal{I}_{\textrm{bc}}(\nu, \Delta) =&\mathcal{F}\left[\left(I(t,\Delta)-\lvert \alpha_{t} \rvert^2 I_a(t) \right. \right.\nonumber \\
& \left.\left. - \lvert \alpha_a\rvert^2 I_t(t) \right)\Theta(t-t_1)\right]\,, \label{eq.ResIntensity}
\end{align}
where $I(t,\Delta)$ is the experimentally accessible time- and frequency-resolved intensity, $I_a(t)$ and $I_t(t)$ are the individual time-spectra of analyzer and target, respectively, and $\alpha_a$, $\alpha_t$ the prefactors describing electronic absorption. Compared to the discussion of the \ffc spectrum in the Sec.~\ref{sec:theory}, these two terms correspond to the single-stage contributions $\hat{T}_t \star \hat{T}_t$ and $\hat{S}_{a} \star \hat{S}_a$ in the first line of Eq.~(\ref{eq.IntOff}) In the following discussion we will refer to $\mathcal{I}_{\textrm{bc}}(\nu, \Delta)$ as the background-corrected \ffc spectrum.

In practice, this quantity can be measured in different ways. One approach is to measure $I(t,\Delta)$, $I_a(t)$ and $I_t(t)$ separately. One then has to consider possible variations in the total number of resonant incident photons contributing to the different spectra in order to perform the subtraction. Alternatively, one can measure $I(t,\Delta)$ only, and then determine the $\Delta$-independent background to be subtracted directly in the \ffc spectrum in regions far away from the diagonal structures. This latter approach will be discussed in more detail in Sec.~\ref{sec.PhaseCycle}.

As discussed in Sec.~\ref{sec:large-det}, the background correction in Eq.~(\ref{eq.ResIntensity}) removes the horizontal lines and low-frequency background as shown in Fig.~\ref{fig.Unravel}. Further, the Heaviside $\Theta$-function serves to exclude the first few nanoseconds of the recorded intensity after the nuclear excitation, mainly for two reasons: First, from a theoretical point of view, it is desirable to exclude the contribution from the prompt pulse from the analysis since it is orders of magnitude more intense than the nuclear resonant response and thus leads to a large background in the Fourier spectrum. This lower cutoff can, in principle, be set during data evaluation if the spectrum has been measured reliably at all times. However, the second reason for rejecting the intensity at early times is the fact that the prompt radiation leads to a detector overload during the first few nanoseconds after pulse arrival. Thus, typically the first  few ns are not available for data evaluation.  Throughout the analysis, we will first consider the ideal case with all times available for data analysis and then discuss the effect of a finite measurement time range in Sec.~\ref{subsec.Gating}.

Regarding the discussion of the \ffc spectrum and diagonal structure around Eq.~(\ref{eq.IntOff}) and Eq.~(\ref{eq.OffResult}), the background removal via the time-gating $\Theta(t-t_1)$ in Eq.(\ref{eq.ResIntensity}) removes the off-resonant background proportional to $\lvert \alpha_t \rvert^2$ in the second line of Eq.~(\ref{eq.IntOff}) (for details, the reader is referred to the discussion in Sec.~\ref{sec:large-det}) and thus the background-corrected \ffc spectrum is expected to be described well by Eq.~(\ref{eq.OffResult}) in the large analyzer-target detuning limit, which will be exploited to reconstruct the complex-valued nuclear resonant target response in Sec.~\ref{sec:nuc-response}.

\begin{table*}[ht]
\caption{\label{tab.HyperfineRes}%
Magnetic hyperfine splittings of ${}^{57}$Fe extracted using the  line fits to the off-resonant \ffc spectrum's diagonal structures. The results compare results with and without optimization of the fit range, and corresponding data for two different time gatings. The first column indicates reference values obtained from a calculation using {\sc pynuss}. All results are given in units of the single-nucleus line-width $\gamma$.
}
\begin{ruledtabular}
\begin{tabular}{lllll}
\multicolumn{1}{c}{\textrm{\textsc{pynuss} Ref.}}&
\multicolumn{1}{c}{\textrm{Off-resonant}}&
\multicolumn{1}{c}{\textrm{Off-resonant (opt.)}}&
\multicolumn{1}{c}{\textrm{Gating ($20$ns)}}&
\multicolumn{1}{c}{\textrm{Gating ($40$ns)}}\\
\colrule
8.7596 & 8.652(25) & 8.719(12) & 8.83(32) & 9.32(39) \\
32.0605 & 31.956(27)& 32.053(17) & 32.35 (51) & 32.32(52) \\
55.3406 & 55.159(42) & 55.267(15) & 56.00(55) & 55.1(1.1) \\
\end{tabular}
\end{ruledtabular}
\end{table*}

\subsection{\label{sec.expar}Extracting spectral parameters via linear fits to the diagonal structure}

We start by performing linear fits to the diagonal structure, which we then extrapolate towards the crossing points with the $\nu$- or $\Delta$-axes, in order to extract spectral information on the target such as resonance energies. In spectroscopic applications of Nuclear Resonant Scattering, the spectra often feature multiple splitted, broadened and shifted resonances, e.g., due to magnetic or electric hyperfine fields. In such a multi-resonance case, the condition Eq.~(\ref{eq.diag}) generalizes to 
\begin{align}
\nu_j(\Delta) = \pm(\Delta - \omega_{\textrm{hf}}^{(j)}+\omega_a) \label{eq.multiline}
\end{align}
with slope $\pm 1$ and points of intersection $ (\omega_a-\omega_{\textrm{hf}}^{(j)})$ with the $\nu$-axis for each transition resonance frequency $\omega_{\textrm{hf}}^{(j)}$ separately. In order to extract these resonance frequencies via linear fits to the diagonal structures in the \ffc spectrum, one has to determine points along the diagonals through which the lines should be fitted. In the following, we discuss two approaches to perform such linear fits, based on horizontal or vertical sections through the \ffc spectrum. Note that in both cases, each resonance frequency $\omega_{\textrm{hf}}^{(j)}$ is obtained via a fit of a diagonal line with the offset as the single free parameter, which combines data over a broad range of detunings values, thereby reducing the detrimental effect of statistical or systematic uncertainties. These fits do not require prior knowledge about the target systems, and in this sense are model-independent. This suggests the diagonal analysis of \ffc spectra as a versatile spectroscopic technique at pulsed x-ray sources.

As a specific example, we extract the magnetic hyperfine splittings of $^{57}\textrm{Fe}$ with internal hyperfine field $B = 33.3$T from the FFC spectra. As a reference to compare the linear fit results to, we determine the three positive hyperfine transition frequencies as peak maxima of the nuclear resonant target spectrum calculated with \textsc{pynuss} for a $0.3$µm thin target foil. The results are tabulated as a function of the single-nucleus line-width $\gamma$ in Table~\ref{tab.HyperfineRes} in the column ``\textsc{pynuss} Ref.''.

Figure~\ref{fig.extract-fit}(a) shows the real part of the background-corrected \ffc spectrum of a $2\mu$m thick $\alpha\!-\!{}^{57}\textrm{Fe}$ foil irradiated by linearly polarized light in forward scattering geometry as given in Eq.~(\ref{eq.ResIntensity}). The magnetic hyperfine field is oriented perpendicular to the beam propagation direction and tilted by an angle of $45$ degrees with respect to the beam polarization axis to ensure that all hyperfine transitions are adressed during the scattering process. In computing the \ffc spectrum, we first consider the ideal case, by including intensities from times slightly after the x-ray excitation in order to suppress the prompt non-resonant contribution, up to measurement times larger than $10\gamma^{-1}$ to ensure high resolution along the $\nu$-axis. The effect of more restricted measurement time intervals will be discussed in Sec.~\ref{subsec.Gating} below.
Further, we choose the real part of the (complex-valued) \ffc spectrum for the peak evaluation since it is more symmetric and spectrally narrower than its absolute value, as the imaginary part features a broad asymmetric line shape.

\subsubsection{Vertical cuts along the $\nu$-axis}
\begin{figure}[t]
\includegraphics[width = \columnwidth]{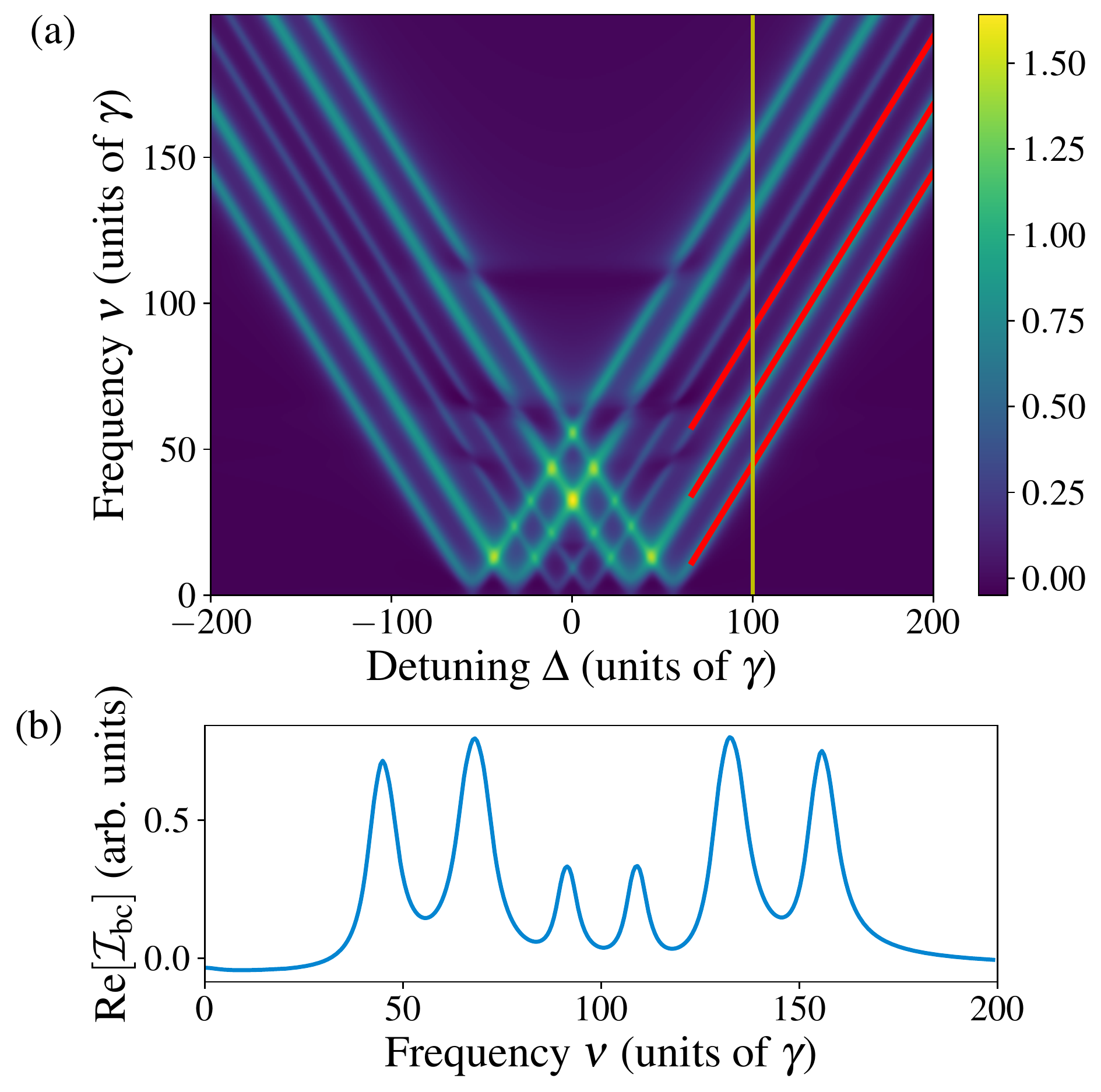}

\caption{Extraction of spectral parameters via linear fits to the diagonal structure. (a) shows the real part of the \ffc spectrum for a $2\mu$m thick $\alpha$-iron enriched in $^{57}\textrm{Fe}$ probed in Nuclear Forward Scattering (see main text for details). The analyzer is a stainless steel foil as described in the main text. The diagonal structure is clearly visible. The line fits used to extract the resonance positions are marked in red. (b) shows a vertical cut through (a) at a detuning $\Delta = 100\gamma$. This detuning is indicated by the yellow vertical line in (a). The six resonances of the magnetically split $\alpha$-Fe spectrum can clearly be distinguished. \label{fig.extract-fit}}
\end{figure}

The first approach employs vertical cuts along the $\nu$-axis through the \ffc spectrum. As an example, we analyze diagonals with positive slope. In order to avoid spectral overlap with diagonals of negative slope, we restrict the analysis to  detunings $\Delta$ which are larger in magnitude than the outermost crossing point of the diagonals with the $\nu=0$ axis, as indicated by the red diagonal lines in Fig.~\ref{fig.extract-fit}(a). For each such $\Delta$, we analyze a vertical cut through the diagonal structure in the \ffc spectrum,  as exemplified by the yellow line in Fig.~\ref{fig.extract-fit}(a). The section corresponding to the yellow line in (a) is shown in (b), which shows that the maxima of the different diagonals are clearly resolved.  The respective lines formed by these maxima across different detunings are then linearly fitted  with a fixed slope of $1$ and offset $b$. The detuning range considered for the linear fits is indicated by the red lines in Fig.\ref{fig.extract-fit}(a), which represent the result of the linear fits. The fit parameters then allow one to determine the transition frequencies $\omega_{\textrm{hf}}^{(j)}$ relative to the resonance frequency $\omega_a$ of the analyzer via Eq.~(\ref{eq.multiline}).

The results of this analysis for the three positive-valued hyperfine splittings $\Delta \omega$ together with the corresponding fit errors are given in Table~\ref{tab.HyperfineRes} in the column ``Off-resonant''. In order to check for residual effects of  resonant contributions on the diagonal structure at small detunings, we further repeated the above analysis for different $\Delta$ fit ranges. To this end, we evaluated 20 fits $j\in\{1, 2, \dots, 20\}$ in which the first $5j$ datapoints (stepsize $\delta \Delta = 0.5 \gamma$) with the lowest detunings $\Delta$ were excluded. Out of those 20 fits, we then chose the one with the lowest fit error as the optimized result. The results of this procedure are shown in column ``Off-resonant (opt.)'' in Table~\ref{tab.HyperfineRes}. The comparison of the unoptimized and optimized off-resonant fit results with the corresponding theory reference values shows a good agreement  within a 2\% margin of the relative error which demonstrates the feasibility of spectral parameter determination from the diagonal structure. Further, the optimized result yields lower fit errors and more accurate results indicating that there indeed is a residual effect of resonant (e.g., radiative coupling) effects or the negative-slope diagonal branches on the positive-slope diagonal structure close to $\Delta = 0$. 

\subsubsection{Horizontal cuts along the $\Delta$-axis\label{sec:hor}}

Alternatively, an analogous  second approach based on horizontal cuts along the $\Delta$-axis is possible. This approach can be advantageous in case of finite measurement time range for the time-dependent intensity imposed by experimental constraints such as the pulse structure of the x-ray source. The reason is that this time range determines the frequency resolution along the $\nu$ axis in the \ffc spectrum. If this resolution becomes too coarse due to the experimental limitations, then an accurate determination of the maxima along the vertical sections may be challenging. An example for this will be discussed in Sec.~\ref{subsec.Gating}. In contrast, the resolution along the horizontal $\Delta$-axis is determined by the M{\"o}ssbauer drive, and can be chosen  independent of the pulsed x-ray source characteristics.
For an analysis along the $\Delta$-axis, smaller detunings $\Delta$ should be avoided since they may be perturbed by resonant effects such as the radiative coupling of analyzer and target. Further, also the tails of the negative-slope diagonal structure branch may lead to a slight asymmetry of the positive-slope branch and vice versa at low $\Delta$. In practice, it is possible to perform both the horizontal and the vertical cut analysis on the same data set, and to compare the results for a consistency check.

Summarizing this part on extracting spectral parameters, we conclude that the results of the analysis underpin the theoretical explanation of the diagonal structure given in Sec.~\ref{sec.Correla} and provide us with an intuitive and simple way of determining the spectral structure of nuclear targets from \ffc spectra.
 
\subsection{Extracting the resonant target response function via sections through the diagonal structure \label{sec:nuc-response}} 

In this Subsection, we consider the second method to extract information from the diagonal structure in the \ffc spectrum, by making use of the fact that horizontal or vertical cuts through the diagonals in the limit of large analyzer-target detuning provide access to the complex-valued resonant target response function. In the large analyzer-target detuning and thin-analyzer limit, this follows from Eq.~(\ref{eq.trideal}), which states that the \ffc spectrum is expected to be essentially proportional to the desired nuclear resonant part of the target response. For thicker analyzers, this result generalizes to Eq.~(\ref{eq.OffResult}), which shows that then the \ffc spectrum is determined by the cross-correlation between nuclear resonant part of the target response and analyzer response.

\begin{figure}
 \includegraphics[width=\columnwidth]{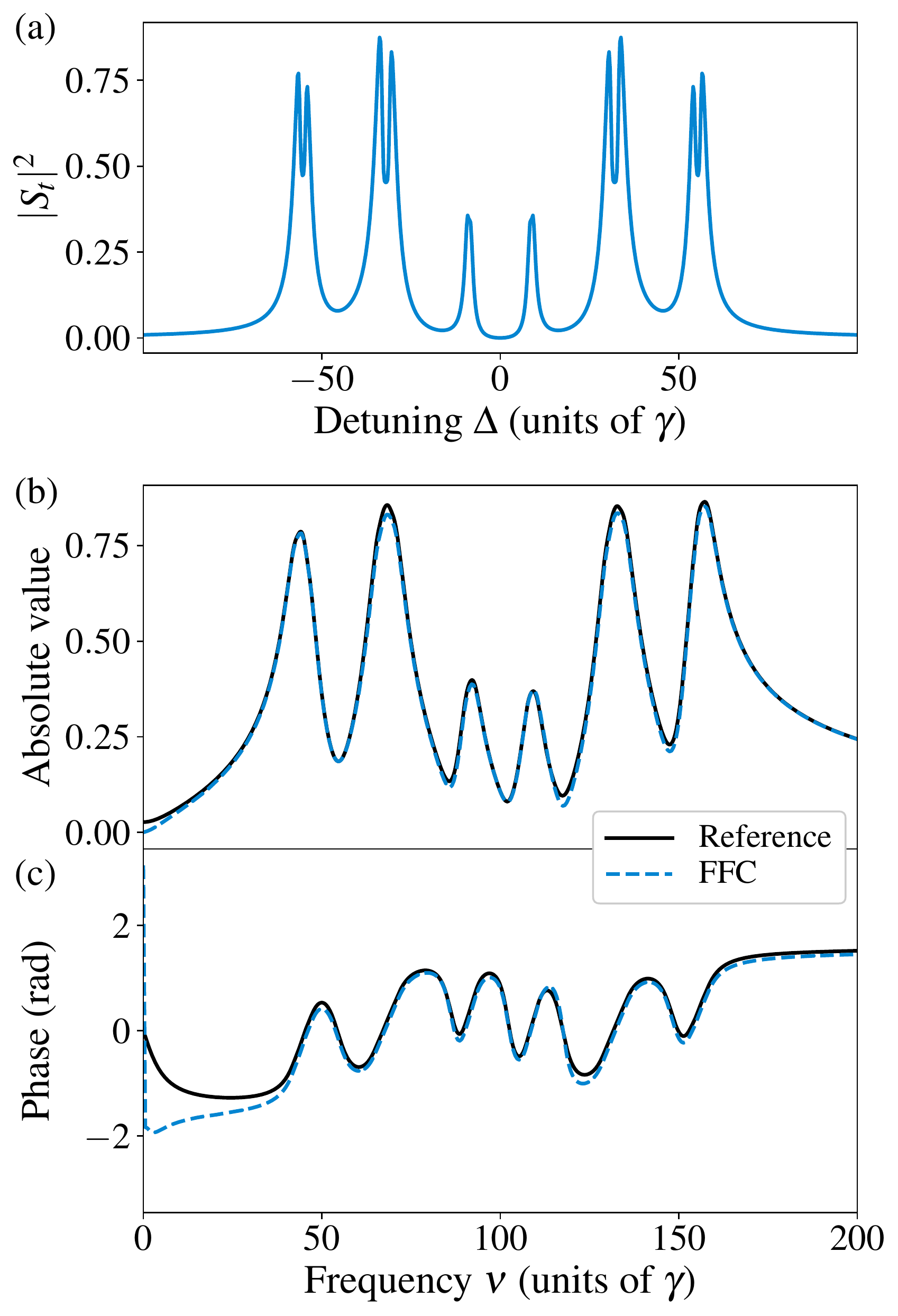}
 \caption{\label{fig.reconstruction}Extraction of the target response function. Results are shown for a target and analyzer configuration as in Fig.~\ref{fig.extract-fit}. (a) shows the true nuclear resonant target spectrum without analyzer as a reference, evaluated using the {\sc pynuss} software package.  (b) depicts the absolute value of a vertical cut at $\Delta = 100\gamma$ through the \ffc spectrum (dashed blue line). For comparison, the absolute value of the reference nuclear resonant target response correlated with the analyzer response according to eq.(\ref{eq.OffResult}) is shown as solid black line. (c) shows the corresponding phase of the two quantities in (b).}
\end{figure}

\subsubsection{\label{extract-nfs}Extraction of the complex-valued resonant target response function in nuclear forward scattering} 

As a first example, we consider the same target and analyzer configuration as in Sec.~\ref{sec.expar}. To this end,
as the reference, we calculate the nuclear resonant target response $\hat S_t(\Delta)$ alone using {\sc pynuss}. Figure~\ref{fig.reconstruction}(a) shows the desired energy spectrum $|\hat S_t(\Delta)|^2$ obtained with this calculation.

Our goal then is to determine the complex-valued nuclear target response  $\hat S_t(\Delta)$ itself from the experimentally accessible \ffc spectrum which, in turn, is complex-valued as the Fourier transform of the time- and frequency-resolved intensity. Following Eq.~(\ref{eq.OffResult}), in the large target-analyzer detuning case, the complex-valued \ffc spectrum corresponds to $\hat S_t(\Delta)$, cross-correlated with the analyzer response $\hat{S}_a(\Delta)$. In order to verify this prediction, we show the absolute value of a vertical cut through the \ffc spectrum at $\Delta = 100\gamma$ as the dashed blue line in Fig.~\ref{fig.reconstruction}(b). The black solid line shows the absolute value of the cross-correlated nuclear target response $\hat{S}_t(\Delta)$, according to Eq.~(\ref{eq.OffResult}), obtained from the reference calculation using {\sc pynuss}. It can be seen that the two curves agree very well.

The corresponding results for the phase of the cross-correlated $\hat S_t(\Delta)$ are shown in Fig.~\ref{fig.reconstruction}(c). Again, the two curves agree well, except for lower frequencies $\nu$, where the phase of the \ffc spectrum deviates from the reference calculation.   This deviation can be attributed to resonant couplings between target and analyzer which lead to a low-frequency background at lower $\nu$. It shows that resonant effects do also affect the spectral shape of vertical sections through the \ffc spectrum and suggests that small values of $\nu$ should be avoided as well. The influence of this defect can be reduced by choosing larger detunings $\Delta$ for the vertical cut, such that the relevant part of the target spectrum moves away from lower $\nu$ values.

The effect of the cross-correlation of the desired target response with the analyzer response in the \ffc spectrum can be seen in the comparison of Fig.~\ref{fig.reconstruction}(a) and (b). The resonances in the spectrum in Fig.~\ref{fig.reconstruction}(a) exhibit so-called double-hump profiles, which is a well-known effect in thicker targets (see, e.g., \cite{Smirnov1999}). In contrast, these effects are not visible in (b) due to the cross-correlation with the analyzer response, which broadens the observed spectrum.

Overall, we find that the \ffc spectrum indeed provides access to the complex-valued nuclear resonant target response function. The broadening due to the cross-correlation with the analyzer response again highlights the usefulness of thin analyzers, which allow one to resolve target spectra with higher spectral resolution.

\begin{figure}[t]
\includegraphics[width=\columnwidth]{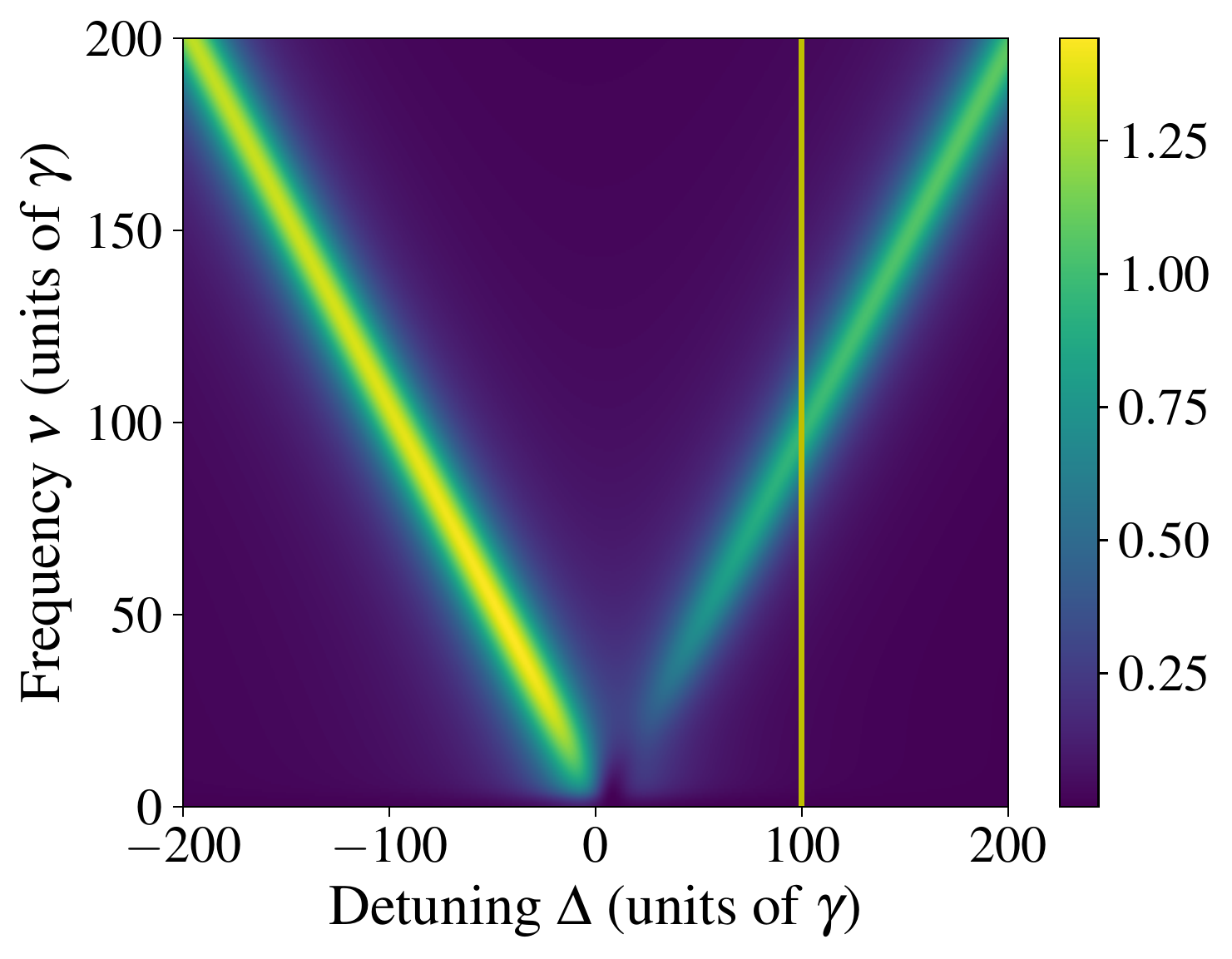}
\caption{Extraction of the complex-valued target response function in reflection for M\"ossbauer nuclei embedded in a thin-film cavity. The Figure shows the absolute value squared of the background-corrected \ffc spectrum as given in Eq.~(\ref{eq.ResIntensity}) with the intensity at time zero set to zero. The cavity structure is 2.2nm Pt/10nm C/0.6nm $^{57}\textrm{Fe}$/10nm C/ Pt (layer structure from the top to the bottom), and the probing x-rays impinge at a fixed incidence angle $\theta = 2.8$mrad. In comparison to Fig.~\ref{fig.extract-fit}(a), the single-line nature of the nuclear response is clearly visible, as well as an asymmetry of the spectrum around $\Delta = 0$. The yellow vertical line indicates the cut at $\Delta=100\gamma$ used in the further analysis. \label{fig.cavity}}
\end{figure}

\subsubsection{Extraction of the complex-valued resonant target response function in reflection from x-ray cavities \label{subsec.Cav}}

As a second example, we discuss the recovery of the complex-valued resonant target response function for M\"ossbauer nuclei embedded in x-ray cavities. Similar to the forward-scattering case in the previous section, the nuclear energy spectra observed in reflection and evaluated by standard methods like the late-time integration are modified by an interference with the electronic response of the cavity structure~\cite{RoehlsbergerBook}. This interference is particularly prominent in the cavity case, since it depends on the x-ray incidence angle around the cavity resonance, and may lead to a full response with an asymmetric Fano profile~\cite{Heeg2015}. For this reason, the extraction of the nuclear response, unperturbed by the electronic cavity response, from the \ffc spectrum is of particular relevance for the field of nuclear quantum optics, as discussed below.

\begin{figure}[t]
\includegraphics[width = 1.0\columnwidth]{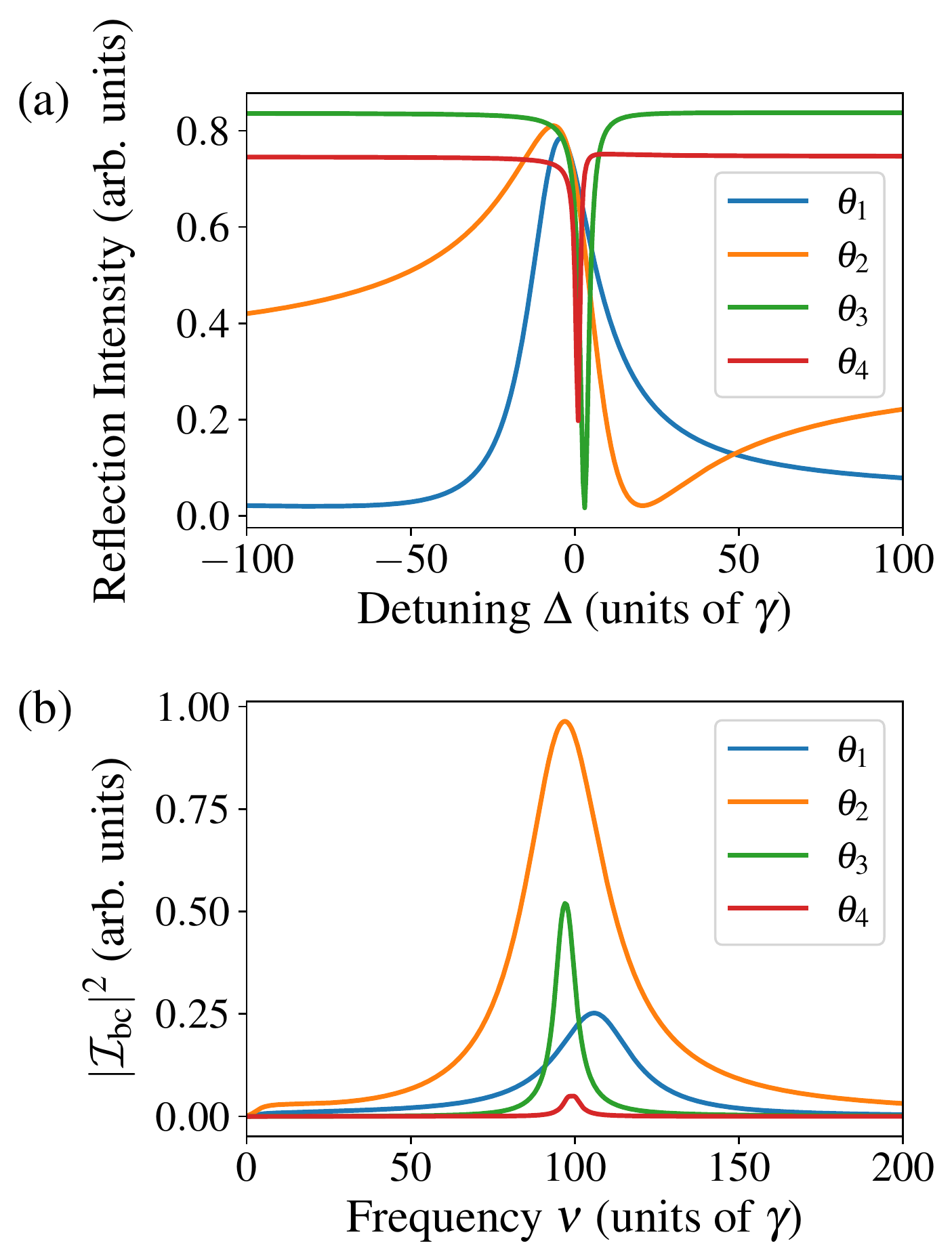}

\caption{\label{fig.CavityPlots} Cavity spectra for different x-ray incidence angles $\theta_1 = 2.75$mrad, $\theta_2=2.8$mrad, $\theta_3=3.0$mrad and $\theta_4=3.3$mrad. (a) shows full reference cavity reflection spectra calculated using the software package {\sc pynuss}. The effect of the electronic scattering on the cavity structure is clearly visible in the off-resonant background varying with the incidence angle, and in the modification of the Lorentzian nuclear response into a Fano line profile. (b) shows the nuclear target response recovered from the \ffc spectrum by vertical sections at $\Delta = 100 \gamma$ (cf. yellow line in Fig.~\ref{fig.cavity}). These spectra are not affected by the electronic scattering on the cavity, and therefore remain of Lorentzian shape for all incidence angles. Therefore,  the dependence of the superradiant line broadening and the resonance energy shift on the incidence angle become clearly visible.
} 
\end{figure}

For the cavity settings, all calculations are performed using the software package {\sc Pynuss}~\cite{Pynuss}, which uses the layer formalism~\cite{RoehlsbergerBook} to  calculate the full (resonant and nonresonant) response, see Sec.~\ref{sec.comp-details}. As our model system, we use a thin-film cavity with layer structure 2.2nm Pt/10nm C/0.6nm $^{57}\textrm{Fe}$/10nm C/ Pt from top to bottom probed by near-resonant x-rays in grazing incidence (cf. Fig.~\ref{fig.Setup}). Here, the last Pt layer is assumed to be sufficiently thick to prevent any transmission through this layer. Due to the low thickness of the Fe layer, magnetic long-range order is absent, and the nuclei do not experience a magnetic hyperfine field. As a result, no magnetic splitting appears in the spectra. Fig.~\ref{fig.cavity} shows the background-corrected \ffc spectrum of such a  cavity at incidence angle $\theta = 2.8$mrad, specifically, the absolute value squared of the quantity given in Eq.~(\ref{eq.ResIntensity}) with the intensity at time zero set to zero.

In contrast to forward scattering, the electronic response of the cavity is different in that it also has a resonance structure, due to the different modes of the cavity. As a result, the magnitude of the off-resonant electronic background observed in reflection varies with the x-rays incidence angle around the cavity resonance, as the variation in the angle effectively scans the incident light frequency across the cavity resonance. On the other hand, a variation of the incidence angle also leads to a relative phase between the electronic and the nuclear response, such that the Lorentzian response of the nuclei appears as Fano line shapes in the total cavity response~\cite{Heeg2015} (for details, see Appendix \ref{sec.ColParams}). Both effects are clearly visible in Fig.~\ref{fig.CavityPlots}(a), which shows reference cavity spectra evaluated  using {\sc pynuss} for different incidence angles. The nuclear response corresponds to the spectral Fano structures around $\Delta = 0$. The off-resonant electronic background becomes visible at large detunings $|\Delta|\gg \gamma$ away from the nuclear resonance.

In Fig.~\ref{fig.CavityPlots}(b), we show the corresponding nuclear target responses reconstructed via vertical cuts through the \ffc spectrum, as illustrated in Fig.~\ref{fig.cavity} for one particular incidence angle. In contrast to the reference spectra in (a), in all cases, the reconstructed resonances are of Lorentzian shape, and the off-resonant background is low. In the following, we will show that this difference to Fig.~\ref{fig.CavityPlots}(a) is due to the fact that the recovery via the \ffc spectrum determines the target response function independent of the electronic scattering contribution. As a result, no off-resonant background contributes, and the nuclear line shape is observed in its original Lorentzian form. We note that this is a qualitative difference to the standard late-time integration method for measuring energy spectra of cavity targets, which does not provide access to the nuclear response alone.

As a first approach to verifying the recovery of the nuclear target response alone, we exploit that the resonances for the various incidence angles in Fig.~\ref{fig.CavityPlots}(b) differ in line-width and center frequency. The variations of the line-width with the incidence angle in Fig.~\ref{fig.CavityPlots}(b) are known in the literature as superradiance $\gamma_s$~\cite{dicke_coherence_1954,gross_superradiance_1982,garraway_dicke_2011,Roehlsberger2010,chumakov_superradiance_2018,HannonTrammell1999,lynch_time_1960,kagan_excitation_1979,guerin_light_2017} and the variations in the resonance position are related to the collective Lamb shift $\Delta_{CLS}$~\cite{Roehlsberger2010,longo2016tailoring,shv,friedberg_frequency_1973,scully_collective_2009,ruostekoski_emergence_2016,keaveney_cooperative_2012,peyrot_collective_2018,roof_observation_2016,bromley_collective_2016}.
They are of particular relevance, since the manipulation of the resonance properties via these collective effects forms the basis for the implementation of more advanced quantum optical level schemes with nuclei~\cite{Roehlsberger2012,Heeg2013Model,Heeg2013SGC,Heeg2015SL,haber_collective_2016,haber_rabi_2017,Lentrodt2020,kong_greens-function_2020,PhysRevA.105.013715,Diekmann2022,yoshida_quantum_2021}.

In the ideal two-level case, the nuclear response comprises a single Lorentzian~\cite{Heeg2013Model,Lentrodt2020,kong_greens-function_2020} which can be characterized by the parameters $\gamma_s$ and $\Delta_{CLS}$ only. In the following, we therefore show that these nuclear parameters can be determined from \ffc spectra, and thereby the nuclear response. To this end, we employ a fit model to extract $\gamma_s$ and $\Delta_{CLS}$ from cuts through the \ffc spectra (details on the fit model are provided in Appendix~\ref{sec.ColParams}). For $\Delta_{CLS}$, we alternatively use a simple determination of the maxima of the cuts through the \ffc spectrum, similar to the resonance determination in Sec.~\ref{extract-nfs}.

\begin{figure}[t]
\includegraphics[width=0.95\columnwidth]{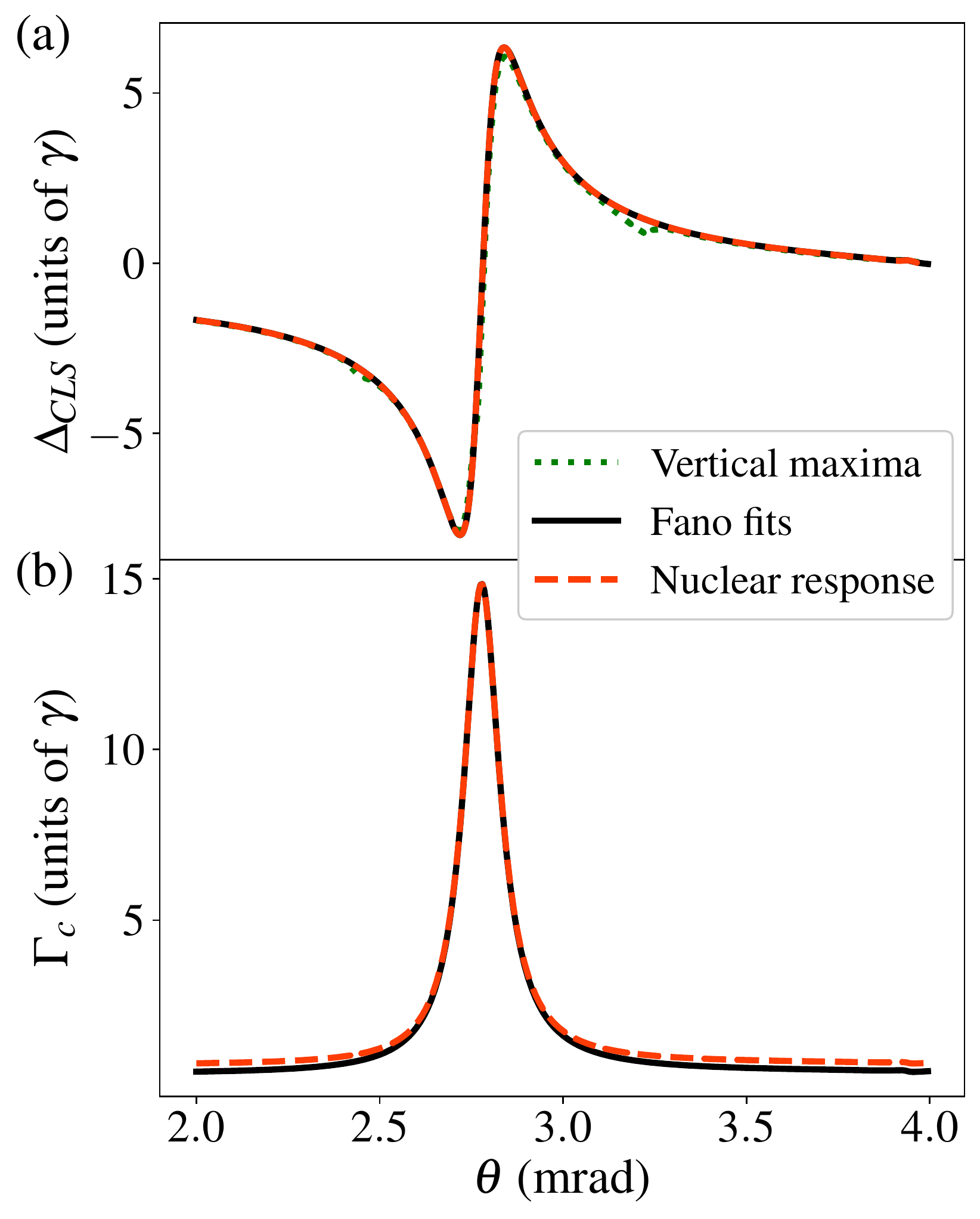}
\caption{Collective nuclear parameters of the two-level system realized by the cavity structure in Fig.~\ref{fig.cavity}. (a) shows the energy shift of the nuclear resonance related to the collective Lamb shift as function of the x-ray incidence angle, and (b) the corresponding superradiantly enhanced decay rate. The red dashed lines are the recovery results obtained from fitting a model to the \ffc spectrum (see main text for details). The green dashed line in (a) is obtained using a simple maximum determination. As a reference, the solid black line shows the parameters obtained via a Fano line fit to the reference cavity spectrum calculated using {\sc pynuss}. \label{fig.CavParams}}
\end{figure}

The results are shown in Fig.~\ref{fig.CavParams}. Panel (a) shows the shift $\Delta_{CLS}$ of the resonance energy as function of the incidence angle $\theta$. The red dashed line depicts the parameters extracted via the fit to the cut through the \ffc spectrum. Results from the simple maxima determination are shown as the green dashed line. In comparison, the black solid line shows the corresponding values obtained via Fano line fits to the reference cavity spectra in Fig.~\ref{fig.CavityPlots}(a) calculated using {\sc pynuss}.  It can be seen that the overall agreement of all curves is good across the entire angular range. At two incidence angles, around $\theta \approx 2.5$mrad and $\theta \approx 3.1$mrad, the simple maxima determination method suffers slightly larger deviations from the other two curves. This deviation can be traced back to uncertainties in the maxima determination because of the residual double-hump spectrum of the analyzer, which becomes relevant since the line-width of the cavity falls below the effective analyzer width at off-resonant angles and thus the shape of the analyzer response dominates the cross-correlation of both response functions.

Fig.~\ref{fig.CavParams}(b) shows corresponding results for the collectively-enhanced total decay rate $\Gamma_c = (\gamma+\gamma_s)/2$ as function of the incidence angle $\theta$. As expected~\cite{HannonTrammell1999,shv,Roehlsberger2010,longo2016tailoring}, the superradiance is highest close to the cavity resonance. Again, the agreement between results extracted from the \ffc spectrum to the reference calculations is good. Towards off-resonant incidence angles, the line-width extracted from the \ffc spectrum saturates to a value slightly larger than that of the reference spectra. The reason for this is the convolution with the analyzer response, which sets a lower limit to the line-width in the \ffc spectrum. In contrast, the resonance position in Fig.~\ref{fig.CavParams}(a) is not affected by the broadening due to the cross-correlation with the analyzer response.


\subsubsection{Determination of the complex-valued nuclear response of an EIT cavity}

Next, we demonstrate that the \ffc spectrum also provides direct access to the complex-valued nuclear response of a more complex cavity structure, using the relevant example of a cavity featuring electromagnetically-induced transparency (EIT)~\cite{fleischhauer_electromagnetically_2005,Roehlsberger2012,Lentrodt2020}. More specifically, we consider a layer structure Pd(1.5nm)/ $\textrm{B}_4\textrm{C}$(49.8nm)/ $^{57}\textrm{Fe}$(0.57nm)/ $\textrm{B}_4\textrm{C}$(97.1nm)/ $^{57}\textrm{Fe}$(0.57nm)/ $\textrm{B}_4\textrm{C}$(35.4nm)/ Pd(43.7nm)/ Si at incidence angle $\theta = 2.28$mrad, as it was discussed in~\cite{Diekmann2022}.

\begin{figure}[t]
\includegraphics[width = 0.95\columnwidth]{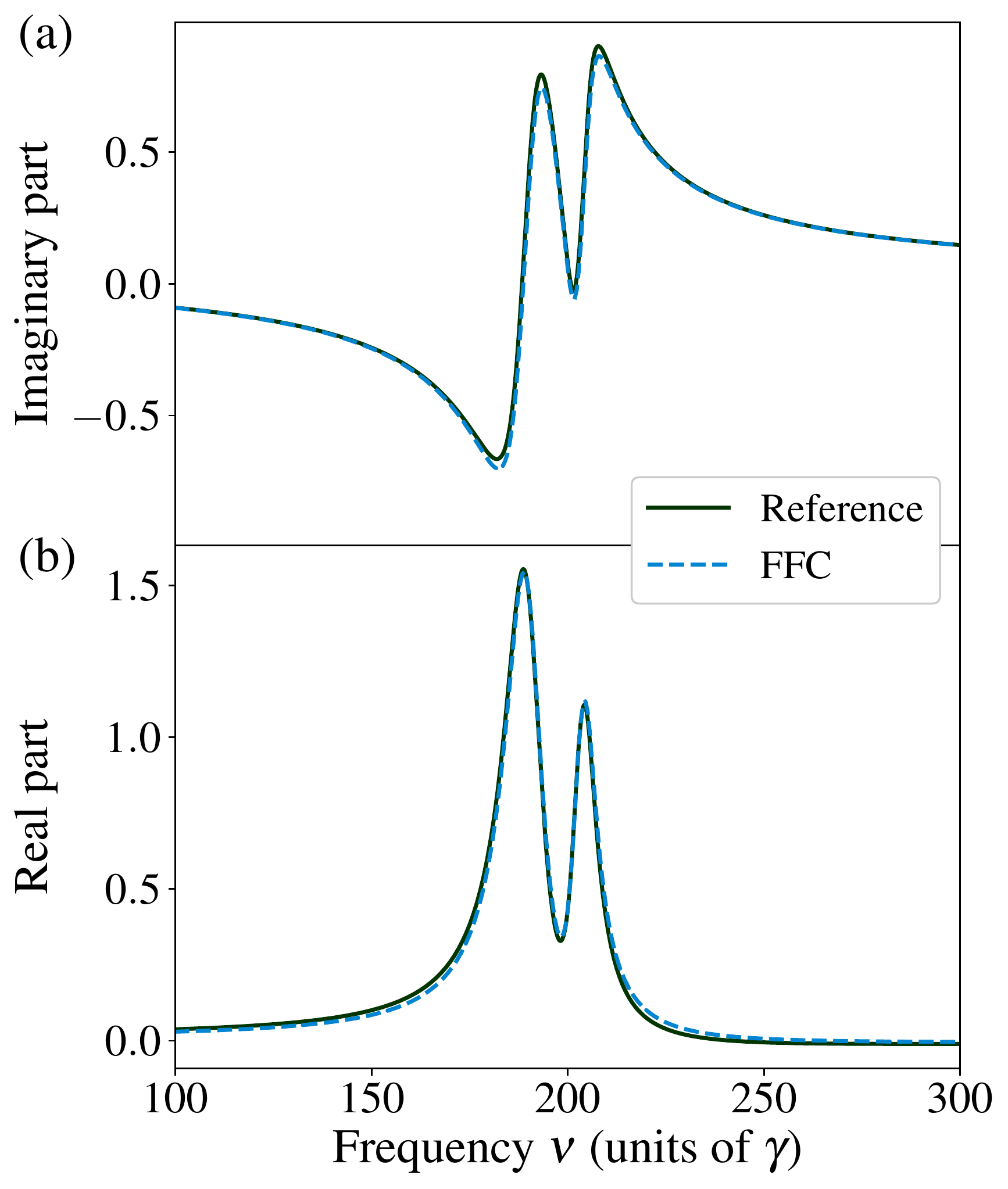}
\caption{Extraction of the nuclear response of a cavity-target probed in reflection. The figure shows the (a) real (b) imaginary parts of a cut through the diagonal structure in the \ffc spectrum along the $\nu$ axis at detuning  $\Delta = -200 \gamma$ (dashed blue curves). The solid black lines show the corresponding cross-correlation of the nuclear resonant responses of cavity and analyzer as given by Eq.~(\ref{eq.OffResult}) as a reference. The dip in the spectra is an electromagnetically induced transparency feature.
The cavity layer structure is Pd(1.5nm)/ $\textrm{B}_4\textrm{C}$(49.8nm)/ $^{57}\textrm{Fe}$(0.57nm)/ $\textrm{B}_4\textrm{C}$(97.1nm)/ $^{57}\textrm{Fe}$(0.57nm)/ $\textrm{B}_4\textrm{C}$(35.4nm)/ Pd(43.7nm)/ Si, and the x-ray incidence angle is  $\theta = 2.28$mrad. \label{fig.EIT}}
\end{figure}

To recover the nuclear resonant target response from the diagonal structure, we consider vertical cuts through the \ffc spectrum at $\Delta = -200 \gamma$, and show the corresponding real and imaginary parts in Fig.~\ref{fig.EIT}. As a reference, the figure further shows the cross-correlation between the resonant target and analyzer responses calculated using  {\sc pynuss}, i.e., it does not contain contributions from the electronic response of the cavity.
As in the forward scattering case, the spectral shape of both real and imaginary part is reproduced well by the \ffc spectrum, and we again find that the \ffc spectrum calculated from the diagonal spectra provides direct access to the complex-valued nuclear target response. Note, that evaluation of the \ffc spectrum at larger detunings $\Delta$ compared to the discussion in Sec.~\ref{extract-nfs} reduces the low-frequency background that impedes the full response reconstruction in the forward scattering case Fig.~\ref{fig.reconstruction}.

To summarize the results of Sec.~\ref{sec:nuc-response}, we showed that the complex-valued target response function can be recovered from the \ffc spectrum in the case of nuclear forward scattering as well as in the case of reflection from a cavity. In the latter case, it is important to note that the method presented here indeed provides access to the complex-valued nuclear target spectrum, independent of the background of and interference with possible electronic scattering channels usually present in the cavity reflection spectra. This is a key difference to other spectroscopy techniques such as the late-time integration, which retrieve the absolute value of the total cavity reflection spectrum. The possibility to access the complex-valued nuclear response alone is of considerable interest for the further development of nuclear cavity QED, as discussed in~\cite{Diekmann2022}, since the spectra corresponding to the nuclear response relate to the quantum optical level scheme governing the nuclear dynamics inside the cavity. Furthermore, from the recovered spectra, also the collectively-modified quantum optical level scheme parameters can be extracted. Therefore, the techniques presented here are expected to fuel further developments in nuclear quantum optics.

\subsection{Finite Gating Times \label{subsec.Gating}}

Both, the theoretical discussion and the numerical results presented so far have assumed the ideal case of all times from arrival of the prompt pulse at $t_1 = 0$ at the detector until a time $t_2 \gg 1/\gamma$, where practically all primary and secondary radiation has decayed, being available for data analysis. However, under realistic experimental conditions the first nanoseconds after pulse arrival can not be used for reliable data evaluation due to detector overload caused by the large intensities of the prompt radiation produced by accelerator-based light sources (cf.\cite{Coussement2000}). On the other hand, repetition rates of these light sources set an upper limit to recording time after pulse arrival before the next bunch hits the target and detector (cf. \cite{RoehlsbergerBook}). To understand the effect of these restrictions on the \ffc spectra, we first reconsider the theoretical derivation of the Fourier-transformed time- and frequency-resolved intensity given in section \ref{sec.Correla} by introducing finite integration boundaries and, subsequently, discuss the applicability of diagonal analysis of magnetic hyperfine splitting in $^{57}\textrm{Fe}$ under these circumstances.

\subsubsection{Theoretical analysis}
For the theoretical discussion, we formally introduce Heaviside $\Theta$-functions excluding the interval $[-t_1,t_1]$ and times above $t_2$ as well as below $-t_2$ to ensure proper symmetry in the Fourier domain. The usual lower cutoff of the integral boundaries in the time-domain necessary for causality reasons is part of the target response, i.e. $T_t(t) \sim \Theta(t)$ (cf. Appendix~\ref{sec.NFSLines}), and gives us this freedom in choosing the form of the gating function at negative times in order to simplify the theoretical analysis. In this case, the Fourier-transformed intensity reads

\begin{align}
\FourierInt{t_1,t_2}  =& E^2_0\int^{\infty}_{-\infty} dt e^{i\nu t} T^*(t)T(t)\Theta(\lvert t\rvert-t_1)\Theta(t_2-\lvert t\rvert) \nonumber \\
=& \frac{E^2_0}{4\pi^2}\Big\{\mathcal{F}\left[\Theta(\lvert t\rvert-t_1)\Theta(t_2-\lvert t\rvert)\right]  \nonumber\\[1ex]
& \qquad\ast \mathcal{F}\left[T^* T\right])\Big\}(\nu) \nonumber \\
 =& \frac{E^2_0}{2\pi^2}\left(\delta_{t_1,t_2} \ast (\hat{T} \star \hat{T})\right)(\nu) \label{eq.FourierIntensity}
\end{align}
where the operator $\mathcal{F}$ denotes the Fourier transform and the $(\ast)$ and $(\star)$ operators the convolution and cross-correlation as defined in Eqs.~(\ref{eq.Convolution}) and (\ref{eq.Correlation}), respectively. Further, we introduced
\begin{align}
\delta_{t_1,t_2}(\nu)&= \frac{1}{2}\mathcal{F}\left[\Theta(\lvert t\rvert-t_1)\Theta(t_2-\lvert t\rvert)\right](\nu) \nonumber \\
&= \int^{t_2}_{t_1}dt \cos(\nu t)= \frac{\sin(\nu t_2)-\sin(\nu t_1)}{\nu} \; . \nonumber
\end{align}
The $\delta_{t_1,t_2}$-function acts as a convolutional filter on the autocorrelation of the response $\hat{T}$ of the complete setup, reduces frequency resolution by virtue of the upper integration boundary $t_2$ and leads to additional oscillations in the final spectrum characteristic of the chosen time-window $[t_1,t_2]$ (cf. \cite{Coussement2000,Herkommer2020}). In principle, these oscillations can be reduced by different means, e.g., experimentally by using crossed polarizer-analyzer setups (cf. \cite{Toellner1995}), by choosing sufficiently late lower cutoff times $t_1$ (late-time integration) \cite{Roehlsberger2010,Heeg2015} (see also Appendix~\ref{app:late}), by smoothing the time window of the lower cutoff during data evaluation \cite{Callens2005}, or by stroboscopic detection techniques~\cite{PhysRevB.65.180404,PhysRevB.67.104423}. Furthermore, event-based data acquisition providing access to time- and frequency-resolved two-dimensional datasets allow one to choose and optimize the integration limits after the experiment throughout the data analysis~\cite{Heeg2017,Heeg2021}. There are also established deconvolution techniques which, however, face the challenge that the filter function  $\delta_{t_1,t_2}$ is zero outside the time range $[t_1,t_2]$.

Finally, we note that the time-gated expression Eq.~(\ref{eq.FourierIntensity}) is consistent with the previous results without time-gating, since
\begin{align}
\delta_{t_1,t_2}(\nu) \xrightarrow[t_2 \to \infty]{t_1 \to 0}\pi\delta(\nu)
\end{align}
in the limit of vanishing time gating, where $\delta(\nu)$ is the Dirac delta distribution. Then, Eq.~(\ref{eq.FourierIntensity}) reduces to Eq.~(\ref{eq.SpecAuto}).

\begin{figure}[t]
\includegraphics[width=0.9\columnwidth]{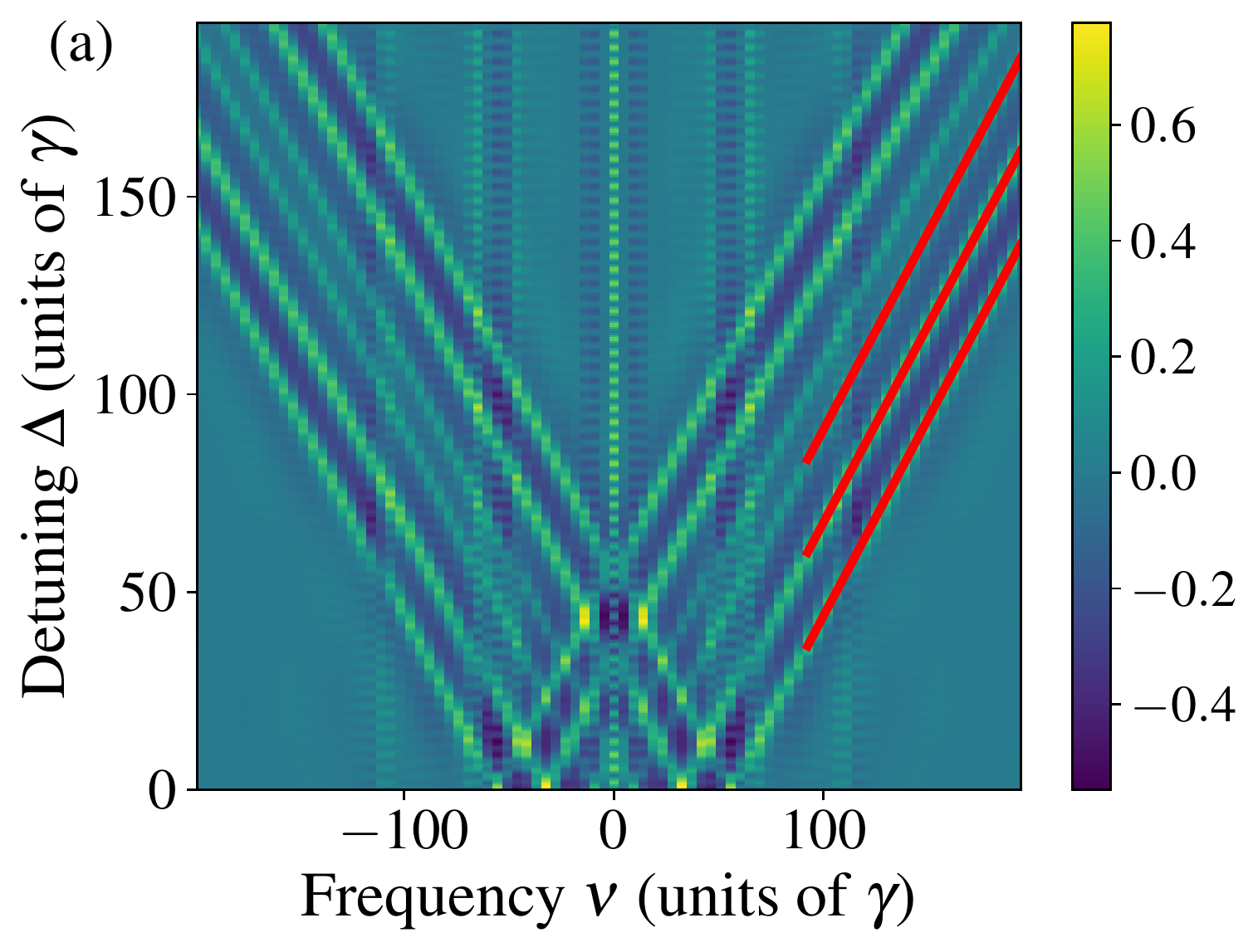}
\includegraphics[width=0.9\columnwidth]{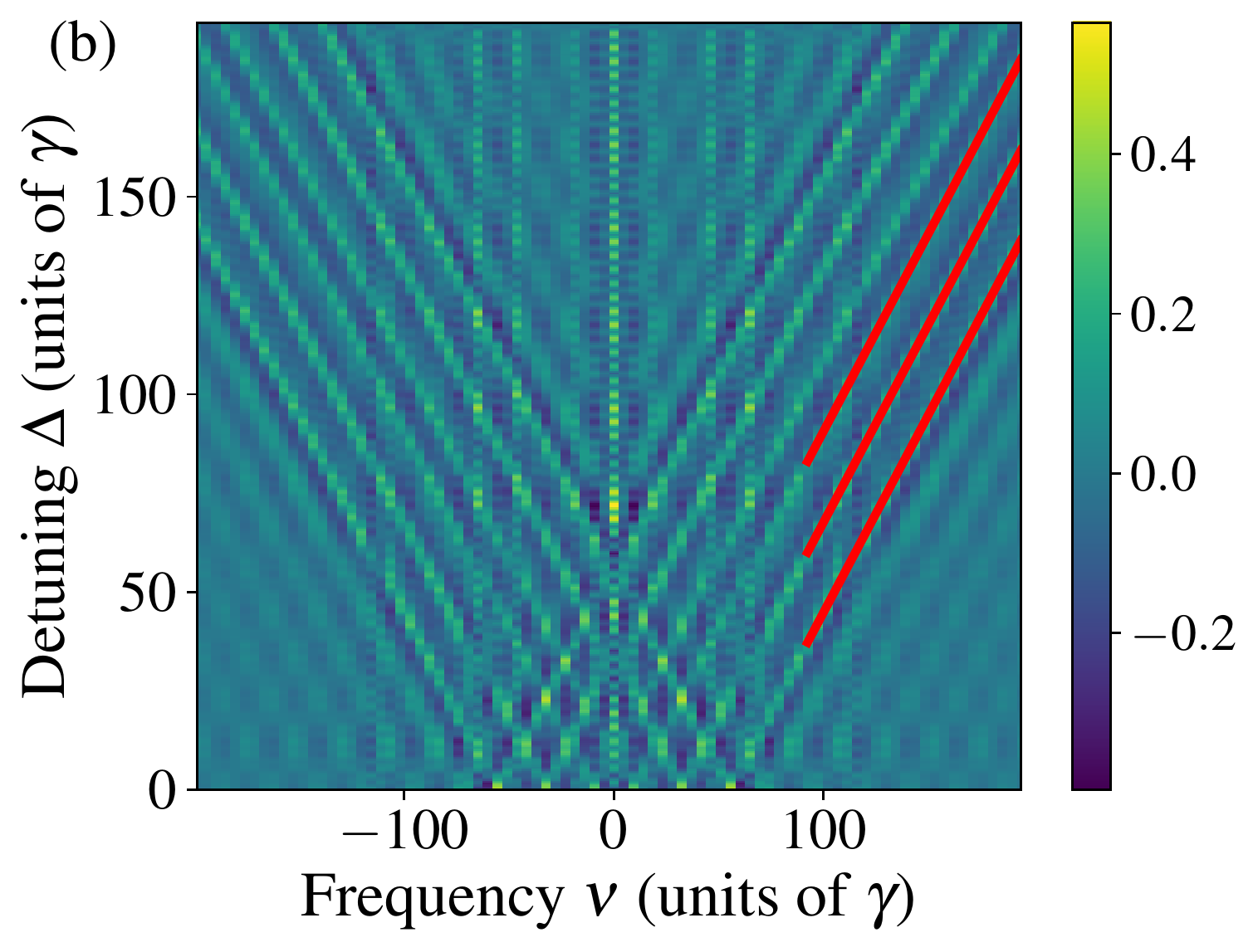}
\caption{\label{fig.Gating}Effect of time gating on the extraction of target parameters from the diagonals in the \ffc spectra. The two panels show \ffc spectra as in Fig.~\ref{fig.extract-fit}, but  with initial time vetos of (a) $20$~ns and (b) $40$~ns. In both cases, only times up to an upper limit of 192~ns are considered. Although the Fourier effects due to the time windows are clearly visible, the data can still be reliably fitted with lines of slope one, as indicated by the red lines.}
\end{figure}

\subsubsection{Effect of temporal gating on the analysis of diagonal spectra}
After having derived the expression for the \ffc spectrum in the presence of gating, we now consider the effect of the gating on the diagonal spectra analysis. For this, we again analyze our example of the magnetic hyperfine splitting in $^{57}\textrm{Fe}$ as described in Section \ref{sec.expar}, using spectra calculated according to Eq.~(\ref{eq.ResIntensity}) with lower gating times $t_1 = 20$ns, $40$ns and upper integration time $t_2 = 192$ns. The latter upper integration limit corresponds to a typical time windows in experiments performed in the 40-bunch mode at the P01 High Resolution Dynamics Beamline~\cite{Wille_2010} at the synchrotron-radiation facility  PETRA III~\cite{PetraIII}. The results are shown in Fig.~{\ref{fig.Gating}} and Table~\ref{tab.HyperfineRes}. It can be seen that the resolution in Fourier space along the $\nu$ axis is decreased by lowering the upper boundary $t_2$. In addition to the lower resolution introduced via $t_2$, well-known periodic structures distort the spectra, an effect that is stronger in the case of the higher lower integration boundary $t_1 = 40$ns  due to the stronger truncation of the Fourier transform along the time axis. Further, the visibility of the diagonal structures reduces with increasing $t_1$ such that the spectra become difficult to analyze for gating times larger than $t_1 = 40$ns.

To overcome these limitations due to the gating, we applied the analysis procedure introduced in Sec.~\ref{sec.expar} to cuts through the diagonal structure along the $\Delta$ axis, which does not suffer from the reduced resolution, as discussed in Sec.~\ref{sec:hor}. For this reason, the $\Delta-$ and $\nu$-axes are interchanged in Fig.~{\ref{fig.Gating}} as compared to Fig.~\ref{fig.extract-fit}(a).
We found that the line fits marked as red lines in Fig.~\ref{fig.Gating} still could be performed reliably, and the hyperfine splittings obtained from the crossing points with the $\nu$-axis given in Table I under ''Gating ($20$ns)'' and ''Gating ($40$ns)'' are accurate up to relative errors in the few percent range.

We therefore conclude that the peak analysis of the background-corrected \ffc spectra involving vertical or horizontal cuts through the diagonal structure combined with line fits through the series of found peaks allows one to recover the nuclear target parameters via  time- and frequency-resolved Nuclear Resonant Scattering spectra even in the presence of considerable time gatings constraining the measurement. Furthermore, a direct analysis of single cuts allows one to recover the complex-valued nuclear part of the target response, although cross-correlated with the analyzer response. This is particularly interesting for target nuclei embedded in thin-film cavities as the nuclear response provides direct access to the underlying quantum optical level scheme, independent of the usual interference with nonresonant electronic scattering off of the cavity structure. In this sense, our method is  complementary to existing analysis methods which provide access to the complete (resonant and nonresonant) cavity response, thereby rendering the retrieval of the pure nuclear response more difficult.

\section{Analysis of the \ffc spectrum beyond the large target-analyzer detuning limit using phase control \label{sec.PhaseCycle}}

In Sec.~\ref{sec.Correla} we exploited that some of the different scattering pathways contributing to the total detection signal (see Fig.~\ref{fig.Setup}) can be approximately isolated from the others in the large-detuning limit between target and analyzer. This formed the basis for the off-resonant spectral analysis methods discussed in the present work and the previous literature (cf. Sec.~\ref{sec:offres}). However, this approach faces several drawbacks. On the one hand, it fails in spectral regions where resonant and off-resonant contributions to the signal overlap, thereby excluding the possibility to study resonant effects such as radiative couplings between targets and related target-analyzer correlation effects. On the other hand, only a restricted $\Delta$ range of the diagonal spectra can be used for the off-resonant analysis, since these overlapping parts cannot be disentangled close to resonance using the previously described recording and analysis scheme. Motivated by this, in this Section, we discuss a more sophisticated method to separate the total detection signal into its various contributions, which is based on a phase control of the analyzer response. We demonstrate that this method can selectively remove undesired signal contributions and thus allows access to a larger amount of data, thereby improving the stability of the diagonal line fit analysis against residual resonant effects and the reconstruction of the phase-resolved nuclear target response. In the future, it may further be employed to study resonant effects that currently are discarded in the off-resonant analysis.

The basic idea of this approach is to conduct a series of measurements, systematically varying parameters that change the different scattering contributions in a controlled and unique way. Combining the different data sets then allows one to separate the different scattering contributions. For example, in the infrared to ultraviolet regime, coherent phase control of laser pulses is exploited to remove undesired scattering contributions in collinear geometry by so-called phase cycling \cite{Hamm2011,Tan2008,Cho2008,Wituschek2020}. This raises the question, if similar methods could be applied in nuclear forward scattering as well. While the phase between subsequent pulses at state-of-the-art accelerator-based x-ray light sources can not be controlled, the control of the relative phase between the exciting x-ray pulse and the light scattered off of the nuclei has been demonstrated using various approaches. Examples include the rotation of the hyperfine field quantization axis~\cite{Shvyd_ko_1994,Shvydko1994,Shvydko1996,PhysRevLett.103.017401,Liao2012,PhysRevApplied.10.014003}, mechanical displacements of the sample after the excitation~\cite{Helistoe1991,Vagizov2014,PhysRevA.87.013807,Heeg2017,Heeg2021}, or transient changes in the magnetic field strength~\cite{Bocklage2021}. In all cases, a rapid manipulation is applied to the nuclei after the initial excitation has passed the sample, such that it only affects the scattered light. Note that the latter two approaches in principle provide access to  arbitrary relative phases between incident and scattered light.

In the following, we show that such phase control allows one to separate different scattering contributions without having to impose the large detuning limit or to conduct reference measurements of target and analyzer alone. To this end, we denote the controllable relative phase between incident and scattered light in the analyzer as $\phi$.  We consider the ideal case in which the phase is applied near-instantaneously after the initial excitation.
Then, the analyzer response can be written as (cf. Appendix~\ref{sec.TargInv}),
\begin{align}
 T_a(t,\Delta, \phi) &= \alpha_a\left[ \delta(t) + e^{i\phi} S_a(t,\Delta)\right]\,.
\end{align}
It is well-known that the scattering off of two targets in general is not commutative~\cite{PhysRevA.71.023804}.
In the present case, the outgoing field behind both targets takes a different form depending on whether the analyzer is placed before (subscript 1) or behind (subscript 2) the target in the beam propagation direction (for details, see Appendix~\ref{sec.TargInv}):
\begin{align}
E_1(t,\Delta, \phi) =& E_0 \alpha_a \alpha_t\left[\delta(t)+ e^{i\phi}S_a(t,\Delta) \right. \nonumber \\[1ex]
& \left.+ S_t(t)+e^{i\phi}S_{a,t}(t,\Delta)\right]\,, \label{eq.AnaFirst} \\[1ex]
E_2(t,\Delta, \phi) =& E_0\alpha_a\alpha_t\left[\delta(t)+ e^{i\phi}S_a(t,\Delta) \right. \nonumber \\[1ex]
&\left.+ S_t(t)+S_{a,t}(t,\Delta)\right]\,. \label{eq.TargFirst}
\end{align}
In these expressions,
\begin{align}
S_{a,t}(t,\Delta) = \int^{\infty}_{-\infty} d\omega e^{-i\omega t} S_a(\omega, \Delta) S_t(\omega) 
\end{align}
describes the radiative coupling between analyzer and target in the time domain.  In particular, the radiative coupling contribution to the outgoing field depends on the ordering of analyzer and target, since the controllable  phase only affects the coupling term if the analyzer is placed in front of the target (case 1). The reason is that in the reverse ordering (case 2), the analyzer phase change is already completed before the scattered light from the upstream target reaches the analyzer, such that no relative phases appear in the coupling term.

The time- and frequency-resolved intensity for the two orderings then evaluate to
\begin{subequations}
\label{eq.PhaseImprint}
\begin{align}
I_{1}(t,\Delta, \phi) =& I_b+2\IZero\textrm{Re}\left[ e^{i\phi}S^*_tS_a \right. \nonumber \\[1ex]
 & \left.+ e^{i\phi}S^*_tS_{a,t}+S^*_aS_{a,t}\right] \,,\\[1ex]
I_{2}(t,\Delta, \phi) =& I_b+2\IZero\textrm{Re}\left[ e^{i\phi}S^*_tS_a \right. \nonumber \\[1ex]
 & \left.+S^*_tS_{a,t} + e^{i\phi}S^*_aS_{a,t}\right] \,.
\end{align}
\end{subequations}
Here,  $\IZero = E^2_0 \lvert \alpha_a \alpha_t\rvert^2$ as introduced in Sec.~\ref{sec.Correla}. Further, the background intensity is given by
\begin{align}
I_b &= I_p +\IZero \left(\lvert S_t \rvert^2 +\lvert S_a \rvert^2+ \lvert S_{a,t}\rvert^2\right)\,.
\end{align}
$I_p$ comprises the terms containing the $\delta(t)$ function characteristic of the prompt unscattered contributions, describing the incoming radiation as well as interference between the prompt pulse and resonantly scattered radiation at $t=0$. Like the other contributions in $I_b$, it is phase-independent if the prompt radiation has passed before the phase imprint onto the analyzer response has taken place.

The intensities in Eqs.~(\ref{eq.PhaseImprint}) comprise the four contributions $I_b$, $S^*_tS_a$, $S^*_aS_{a,t}$ and $S^*_tS_{a,t}$. By forming suitable sums or differences of two intensities recorded with appropriate phases, any two combinations of the above contributions can be isolated, by suppressing the respective other two.

In the present context, the term $S^*_aS_t$ is of primary interest, as it is the time-domain version of the scattering paths in Eq.~(\ref{eq.OffResult}) which gives rise to the diagonal structure. Therefore, we focus on the combinations
\begin{subequations}
\label{pspectra}
\begin{align}
\mathcal{D}_1(t,\chi)&=I_{1}(\phi = \chi)-I_{1}(\phi = \chi+\pi) \nonumber \\[1ex]
&=4\IZero \textrm{Re}\left[e^{i\chi}\left(S^*_tS_a+S^*_tS_{a,t}\right)\right] \label{eq.term-n1} \\[1ex]
\mathcal{D}_2(t,\chi)&=I_{2}(\phi = \chi)-I_{2}(\phi = \chi+\pi) \nonumber \\[1ex]
&=4\IZero \textrm{Re}\left[e^{i\chi}\left(S^*_tS_a+S^*_aS_{a,t}\right)\right] \label{eq.term-n2} \\[1ex]
\mathcal{S}(t,\chi) &= I_{1}(\phi = \chi+\pi)+I_{2}(\phi= \chi+\pi) \nonumber \\[1ex]
&=2I_b+2\IZero\textrm{Re}\left[-2e^{i\chi}S^*_tS_a\nonumber \right.\\[1ex]
&\left.+(1-e^{i\chi})\left(S^*_a+S^*_t\right)S_{a,t}\right] \label{eq.term-n3} \,.
\end{align}
\end{subequations}
which each isolate $S^*_aS_t$ with one of the three other terms $S^*_aS_{a,t}$, $S^*_tS_{a,t}$ and $I_b$, respectively, if $\chi = 0$ is chosen in $\mathcal{S}$. Note that $\mathcal{D}_1$ and $\mathcal{D}_2$ each involve one particular ordering of target and analyzer along the beam propagation direction, while $\mathcal{S}$ is a combination of both orderings. Further, the terms $S^*_aS_t$ in Eqs.~(\ref{pspectra}) feature a prefactor of two as compared to the individual spectra Eqs.~(\ref{eq.PhaseImprint}). Corresponding to doubled statistics for the desired scattering contribution, this compensates for the larger measurement time required to record two spectra for the sum- and difference evaluations. The additional degree of freedom $\chi$ can be used to perform a tomography by scanning the phase of the desired scattering contribution. Such an analysis, however, is beyond the scope of this paper and a value of $\chi = 0$ is chosen as it recovers the form of the diagonal term $S^*_tS_a$ as discussed in previous sections (cf. e.g. Eq.(\ref{eq.CrossCorr})).

In the following, we perform the \ffc analysis developed in Sec.~\ref{sec.Diagonals} on the detection signals in Eqs.~(\ref{eq.term-n1}-\ref{eq.term-n3}).

\begin{figure}[t]
\includegraphics[width = 0.8\columnwidth]{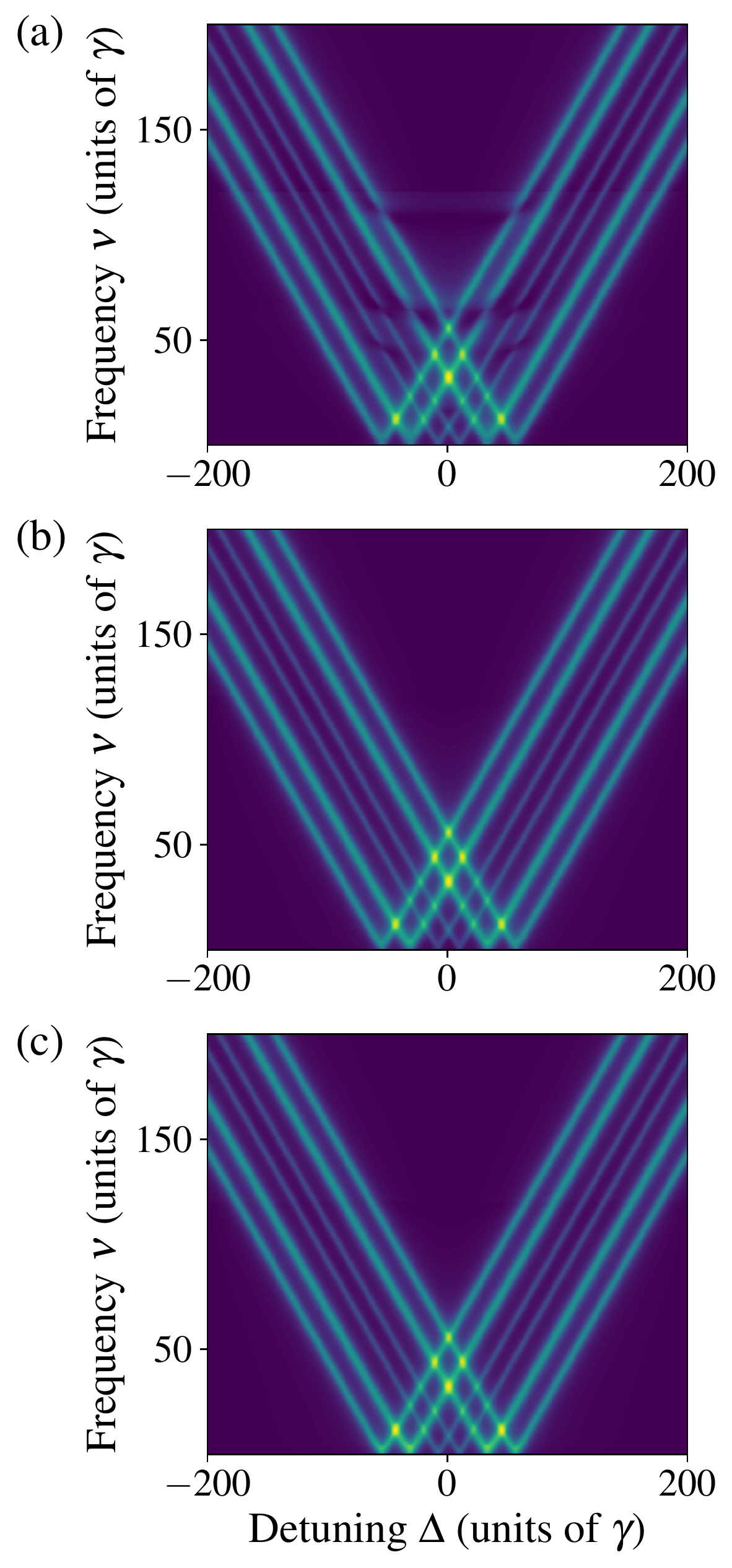}
\caption{\ffc spectra obtained with phase-control of the resonant analyzer response. Each panel shows a combination of two \ffc spectra recorded with different phase settings: (a) shows the phase-combined spectrum $\mathcal{I}_{S_a}$ with  analyzer placed in front of the target. (b) shows the corresponding result $\mathcal{I}_{S_t}$ with analyzer behind the target. (c) shows $\mathcal{I}_{inv}$ after removing the off-resonant background. \label{fig.Combi}}
\end{figure}

\begin{table*}[t]
\caption{\label{tab.HyperfinePhase}%
Magnetic hyperfine splittings of ${}^{57}$Fe extracted from spectra using analyzer phase control. As in Tab.~\ref{tab.HyperfineRes},  line fits to the off-resonant diagonal structures of the \ffc spectrum are used. The different columns compare results for the three phase-combined quantites in Eqs.~(\ref{eq:diffint}), both, with and without optimization of the fit range.  The first column indicates reference values obtained from a calculation using {\sc pynuss}.  All results are given in units of the single-nucleus line-width $\gamma$.
}
\begin{ruledtabular}
\begin{tabular}{llllllll}
\multicolumn{1}{c}{\textrm{\textsc{pynuss} Ref.}}&
\multicolumn{1}{c}{$\mathcal{I}_{S_a}$}&
\multicolumn{1}{c}{$\mathcal{I}_{S_a}$\textrm{ (opt.)}}&
\multicolumn{1}{c}{$\mathcal{I}_{S_t}$}&
\multicolumn{1}{c}{$\mathcal{I}_{S_t}$\textrm{ (opt.)}}&
\multicolumn{1}{c}{$\mathcal{I}_{inv}$}&
\multicolumn{1}{c}{$\mathcal{I}_{inv}$\textrm{ (opt.)}}\\
\colrule
8.7596 & 8.689(23) & 8.748(10) & 8.819(2) & 8.824(2) & 8.854(9)& 8.852(1)\\
32.0605 & 32.032(20) & 32.104(12) & 32.169(7) & 32.193(5) & 32.247(8)& 32.241(4)\\
55.3406 & 55.245(24)  & 55.310(10) & 55.336(13) & 55.374(5) & 55.416(6)& 55.416(6)\\
\end{tabular}
\end{ruledtabular}
\end{table*}

\subsection{Extracting spectral parameters from spectra with phase control \label{sec-phasediag}}

We start with the extraction of spectral parameters as in Sec.~\ref{sec.expar}, by applying line fits of slope one to diagonals formed by the peak maxima of sections along the $\nu$-axis. To allow for a comparison of the different approaches, we again determine the hyperfine splittings of the $2\mu$m thick $\alpha$-iron target with magnetic hyperfine field $B = 33.3$T. As in Sec.~\ref{sec.expar}, the sample is irradiated with linearly polarized light while the hyperfine field is oriented perpendicularly to the beam propagation direction and tilted by $45$ degrees with respect to the beam polarization axis, in order to observe all six transition lines in the spectrum. For the spectral analysis then the real part of the \ffc sum- and difference spectra
\begin{subequations}
\label{eq:diffint}
\begin{align}
\mathcal{I}_{S_a}(\nu, \Delta) &= \mathcal{F}\left[\mathcal{D}_1(t,\chi = 0)\right]  \, , \label{eq.combi-1}\\[1ex]
\mathcal{I}_{S_t}(\nu, \Delta) &=  \mathcal{F}\left[\mathcal{D}_2(t,\chi = 0)\right] \, , \label{eq.combi-2}\\[1ex]
\mathcal{I}_{inv}(\nu, \Delta) &= -\mathcal{F}\left[\mathcal{S}(t,\chi=0) \Theta(t-t_1)\right]-\mathcal{I}_b(\nu_0)  \label{eq.combi-3}
\end{align}
\end{subequations}
are considered. The first spectra do not require additional corrections, since the background contributions are automatically removed by taking the difference of two spectra with phase control in Eqs.~(\ref{eq.term-n1}, \ref{eq.term-n2}). The third spectrum Eq.~(\ref{eq.combi-3}) contains two background corrections. First, the contribution of the initial prompt pulse is suppressed by the step function $\Theta(t-t_1)$. Second, the single-target contributions $\mathcal{I}_b(\nu_0)$ are removed. However, in contrast to the procedure in Eq.~(\ref{eq.ResIntensity}) involving the subtraction of single-target spectra, here, we follow another approach, directly based on the \ffc spectra measured with both targets.  For this, we make use of the fact that the \ffc spectra approximately reduce to the background in regions away from the diagonal structure. Specifically, we determine the background correction from the \ffc contribution at the largest recorded detuning-value $\Delta$ for $\nu$ values up to an upper limit $\nu_0$ well below the diagonal structure. For $\nu$ values above this threshold, we instead approximate the background by the  \ffc contribution at $\Delta = 0$.  This way, the background correction $\mathcal{I}_b(\nu_0)$ is obtained directly from the \ffc spectra, in regions away from the diagonals for the entire $\nu$ range.

For the numerical analysis, we again calculate all spectra using \textsc{pynuss}, including a time-dependent phase shift for the analyzer response which is zero at $t=0$ and $\phi$ for times $t>0$.  The \ffc sum- and difference spectra Eqs.~(\ref{eq.combi-1})-(\ref{eq.combi-3}) considered for the analysis are shown in Fig.~\ref{fig.Combi}. A comparison of plots (a) and (b) shows that the ordering of analyzer and target indeed has an influence: While the diagonal structure is perturbed by residual effects if the analyzer is placed first, corresponding to $\mathcal{I}_{S_a}(\nu,\Delta)$, shown in plot (a), no such perturbations are visible in plot (b) showing the opposite order, corresponding to $\mathcal{I}_{S_t}(\nu, \Delta)$. Like in the case with the target placed first (plot (b)), the background-corrected case $I_{inv}(\nu,\Delta)$ in plot (c) does not show these perturbations.

The retrieved transition frequencies obtained by the diagonal analysis of the phase-combined \ffc spectra Eqs.~(\ref{eq.combi-1})-(\ref{eq.combi-3}), corresponding to spectra recorded with the analyzer first, the target first and upon reversing target order, respectively, are summarized in Table~\ref{tab.HyperfinePhase} in the columns $\mathcal{I}_{S_a}$, $\mathcal{I}_{S_t}$ and $\mathcal{I}_{inv}$ for comparison with the results obtained in Sec.~\ref{sec.expar} displayed in Table \ref{tab.HyperfineRes} without phase control. For the three columns containing the additional ``(opt.)'', again the optimization procedure involving multiple fits with different ranges of detuning data were performed by subsequently excluding low-detuning data points. From the different fit results, the one with the least fit error is chosen. Overall, all results agree well with the respective theoretical reference values within a 2\% margin. We find that the optimization is most effective if the analyzer is placed first ($\mathcal{I}_{S_a}$), which likely is due to the residual perturbations visible in Fig.~\ref{fig.Combi}(a), while the other two results do not change significantly upon optimization. Together with the undistorted appearance of the diagonal structures in these cases in Fig.~\ref{fig.Combi} this implies a larger range of detuning values that can reliably be employed for linear fit analysis compared to the off-resonant case in Sec.~\ref{sec.expar} and Tab.~\ref{tab.HyperfineRes}, since the result with the least fit error is reached very close to resonance already.

\begin{figure}[t]
\includegraphics[width= 0.8\columnwidth]{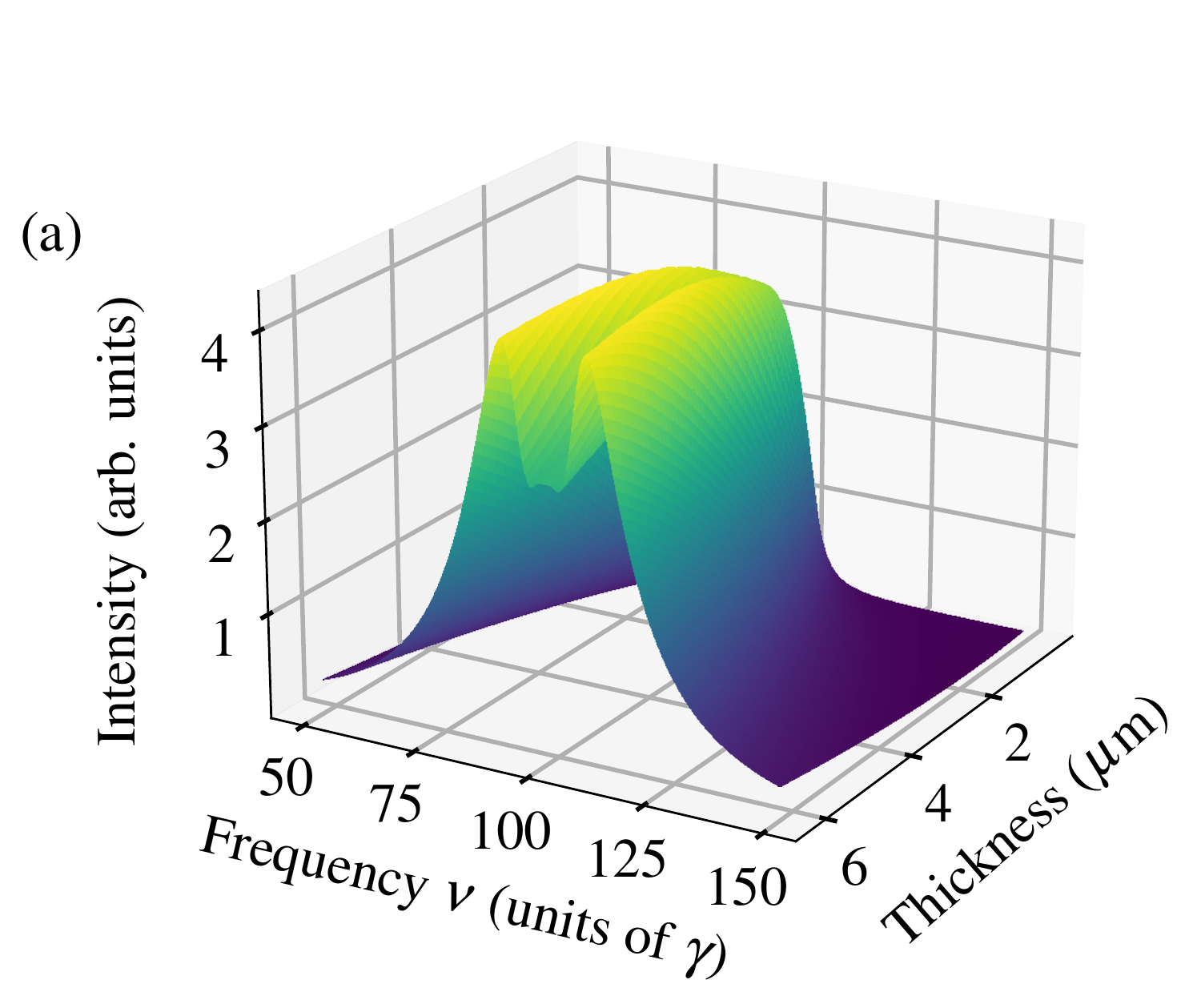}
\includegraphics[width=0.8\columnwidth]{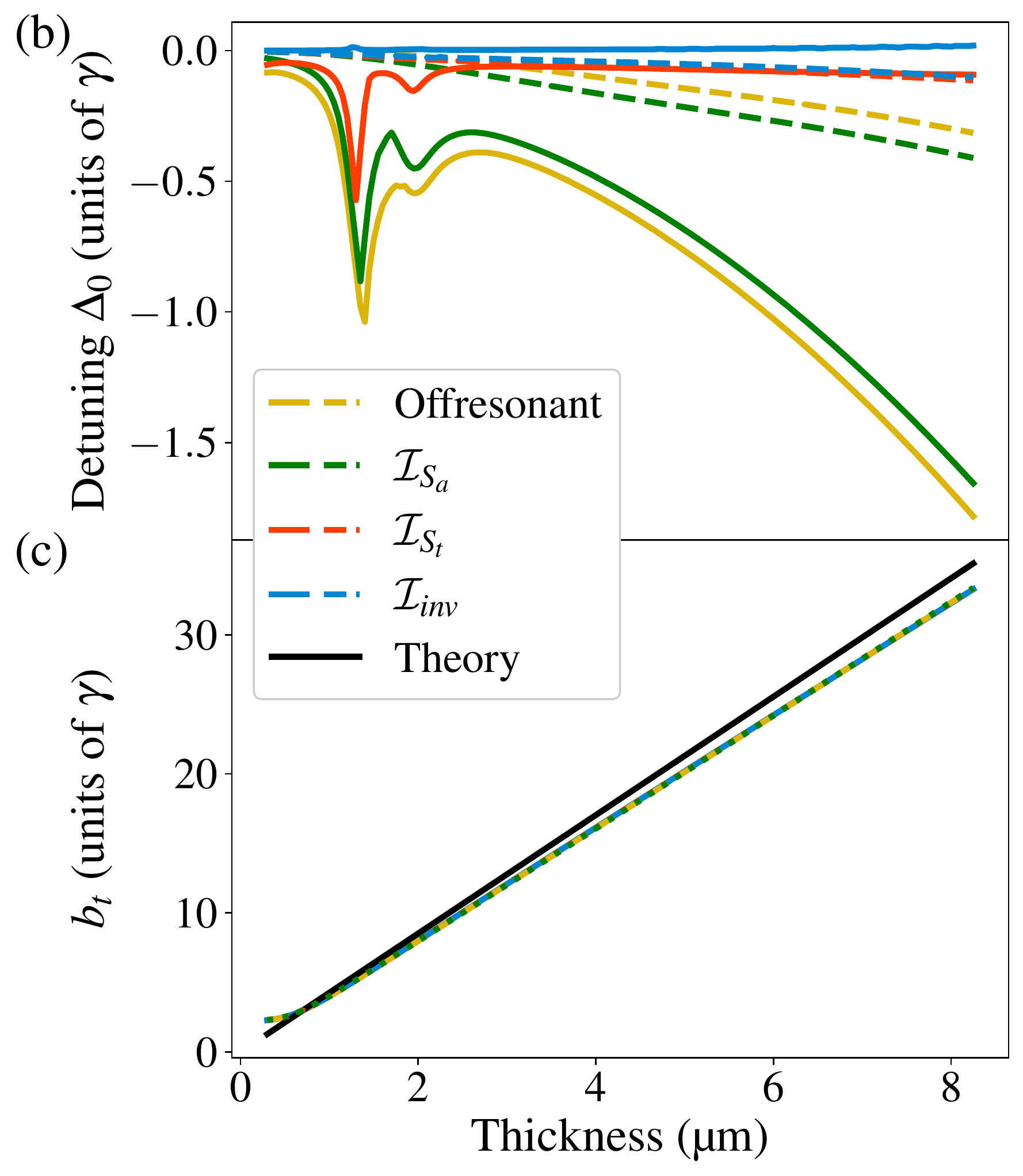}
\caption{Extraction of target parameters as function of target thickness. Panel (a) illustrates how the diagonal structure along the $\nu$-axis at $\Delta = 100 \gamma$  varies with target thickness. For thicker targets, a characteristic ``double-hump'' profile appears.
(b) compares the deviations between the reference transition frequency and the corresponding transition frequencies obtained by linear fits to the diagonals using different methods: The off-resonant method (yellow line), and the phase-difference methods based on $\mathcal{I}_{S_a}$ (green), $\mathcal{I}_{S_t}$ (red) and $\mathcal{I}_{inv}$ (blue). In all cases, the solid lines correspond to results obtained by simple maxima determination while the dashed lines are based on model fits to vertical sections through the diagonal structure, as explained in the main text.
(c) compares the thickness parameters extracted from the different methods with the theoretical reference value (black).  \label{fig.ThicknessScan}}
\end{figure}

\subsection{Effects of the target thickness on the parameter extraction}
For a more systematic comparison of phase-combined and off-resonant \ffc spectra, we analyze the influence of thickness effects on the diagonal structure of a single-line $^{57}\textrm{Fe}$ target (i.e., in the absence of  hyperfine splitting). Fig.~\ref{fig.ThicknessScan}(a) shows the line shape as function of target thickness. It can be seen that for thicker targets, the single line splits into a double-hump profile~\cite{Smirnov1999}, which impedes the simple determination of a single maximum. Therefore, next to the simple fit of a line through the maxima in the \ffc spectra, we further consider a model to fit the line shape of the single-line target, analogous to the approach in Sec.~\ref{subsec.Cav}. We use the model function (for a derivation, see Appendix~\ref{sec.NFSLines})
\begin{align}
f_{\textrm{fit}}(\nu) = A \left[\exp\left(\frac{b_t}{i(\nu-x)-S_{\Gamma}}\right)-1\right] +h \label{eq.ThickModel}
\end{align}
with amplitude $A \sim b_a \IZero$ accounting for electronic absorption and dispersion via $\IZero$ as well as the resonant coupling of the radiation to the analyzer nuclei via $b_a$, $x=\Delta-\omega_t+\omega_a$ denoting the center frequency of each vertical spectrum and $S_{\Gamma} = (\gamma_t+\gamma_a+b_a)/2$ the decay rate of the target broadened by the effective analyzer decay width. The thickness parameters $b_a, b_t$ for both analyzer and target are defined in Appendix~\ref{sec.NFSLines}. Note that, in deriving this formula, we distinguish between analyzer and target line-width $\gamma_a$ and $\gamma_t$ to gain better insight into the functional dependence on both quantities though the analyzed numerical data will continue to use $\gamma_a = \gamma_t = \gamma$.

The exponential form of the fit model Eq.~(\ref{eq.ThickModel}) is a consequence of multiple scattering of photons propagating through the target, also causing the double-hump profile. An additional offset $h$ was added to account for background in the Fourier spectrum due to residual resonant effects. Knowing that the spectral shape of vertical or horizontal sections through the diagonal structure corresponds to a double-hump profile, the simple maxima determination can also be adapted to include the two most prominent maxima of such sections followed by an average over the lines formed by both sets of maxima.\\

Results of this analysis are shown in Fig.~\ref{fig.ThicknessScan}(b). It depicts the recovered target-analyzer detuning $\Delta_0 = \omega_t-\omega_a$ if no Doppler shift is applied as function of target thickness. Since for a $^{57}\textrm{Fe}$ target without hyperfine splitting and a stainless steel analyzer without isomeric shift both transition frequencies are identical ($\omega_t=\omega_a$), the plotted quantity also describes the deviation between the real detuning and the value recovered via diagonal analysis.\\
Eight curves falling into two categories of diagonal analysis methods are compared: The solid curves represent the method introduced in Section \ref{sec.expar} which determines simple maxima of each section along the $\nu$-direction and subsequent linear fits to the resulting diagonals to give an estimate of the target-analyzer detuning $\Delta_0$. The dashed curves use the same linear fit approach to retrieve the target-analyzer detuning but the peak position of each section along the $\nu$-axis is determined using the fit model Eq.~(\ref{eq.ThickModel}) via the parameter $x$. Both of these diagonal analysis methods are applied to four types of \ffc spectra: The yellow curves evaluate the off-resonant spectra discussed in Section \ref{sec.Diagonals} while the other curves result from evaluation of spectra including phase-control, specifically, difference spectra Eq.~(\ref{eq.combi-1})-(\ref{eq.combi-2}) with analyzer (green) and target (orange) placed first in the beam path and the sum spectrum $\mathcal{I}_{inv}$ after background removal (blue). In obtaining these results, the detuning fit range was kept constant as $[25\gamma, 175 \gamma)$.\\

From Fig.~\ref{fig.ThicknessScan}(b) we find that all four simple-maxima evaluations exhibit a maximum deviation from the reference value at around $1.5 \gamma$ which can be attributed to the appearance of the double-hump profile for this target thickness. In contrast, the sum spectrum ($\mathcal{I}_{inv}$, blue curve) is hardly affected. Apart from this, the sum spectrum and the difference spectrum with target placed first ($\mathcal{I}_{S_t}$, orange curve) reproduce the reference value $\Delta_0 = 0$ much better than the other two curves across the entire thickness range, and become almost constant towards larger thicknesses. This feature can be explained by noting that in the corresponding Eqs.~(\ref{eq.combi-2}) and (\ref{eq.combi-3}), the target spectrum is absent except for the desired diagonal contribution. In contrast, Eq.~(\ref{eq.combi-1}) contains further contributions of the target response which distort the diagonal spectrum with increasing target thickness. Hence, to retrieve reliable spectral information for targets thick compared to the analyzer width it appears to be favourable to choose phase-combinations eliminating the term $S^*_t S_{a,t}$ the influence of which becomes significant towards higher target thicknesses.

The results using the model fit approach (dashed lines) all agree well with the expected result of zero throughout the entire thickness range, without perturbation by the double hump at around $1.5\gamma$, since the shape of this double-hump profile is included in the fit model Eq.~(\ref{eq.ThickModel}). However, towards larger target thicknesses, these curves start to deviate more from the reference values than the two better-performing simple-maxima results. We attribute this to the fact that at higher target thickness, both the positive and negative $\nu$ diagonal structures become spectrally broader and overlap stronger with the each other leading to deviations from the fit model Eq.(\ref{eq.ThickModel}). This effect can be reduced by excluding low-frequency data points in the vertical spectra.

Figure~\ref{fig.ThicknessScan}(c) shows corresponding results for the thickness parameter extracted from the four evaluated spectra as function of the target thickness. In all cases, the retrieved thickness parameter are almost identical and agree well with the calculated values for intermediate thicknesses, even though the deviation increases with the target thickness. Also, in the limit of thin targets, a plateau in the recovered thickness parameter develops as the width of the diagonal spectra is bounded from below by the analyzer spectral width.

In summary, we found that the linear fit analysis of spectra obtained using phase control not only allows one to use larger parts of the recorded data than the corresponding off-resonant methods, but also may provide better target parameter recovery results. As a function of target thickness, a comparison between different phase control approaches revealed that it appears most favorable to eliminate the scattering channels containing the term $S^*_t S_{a,t}$. Then, the phase control methods provide a better recovery than the corresponding approaches based on the off-resonant part of the spectrum only. Finally, a more reliable recovery of the spectra as function of target thickness could be achieved using fit models incorporating the known double-hump profile.

\subsection{Extracting the target response function using phase control}

\begin{figure}[t]
\includegraphics[width = 0.9 \columnwidth]{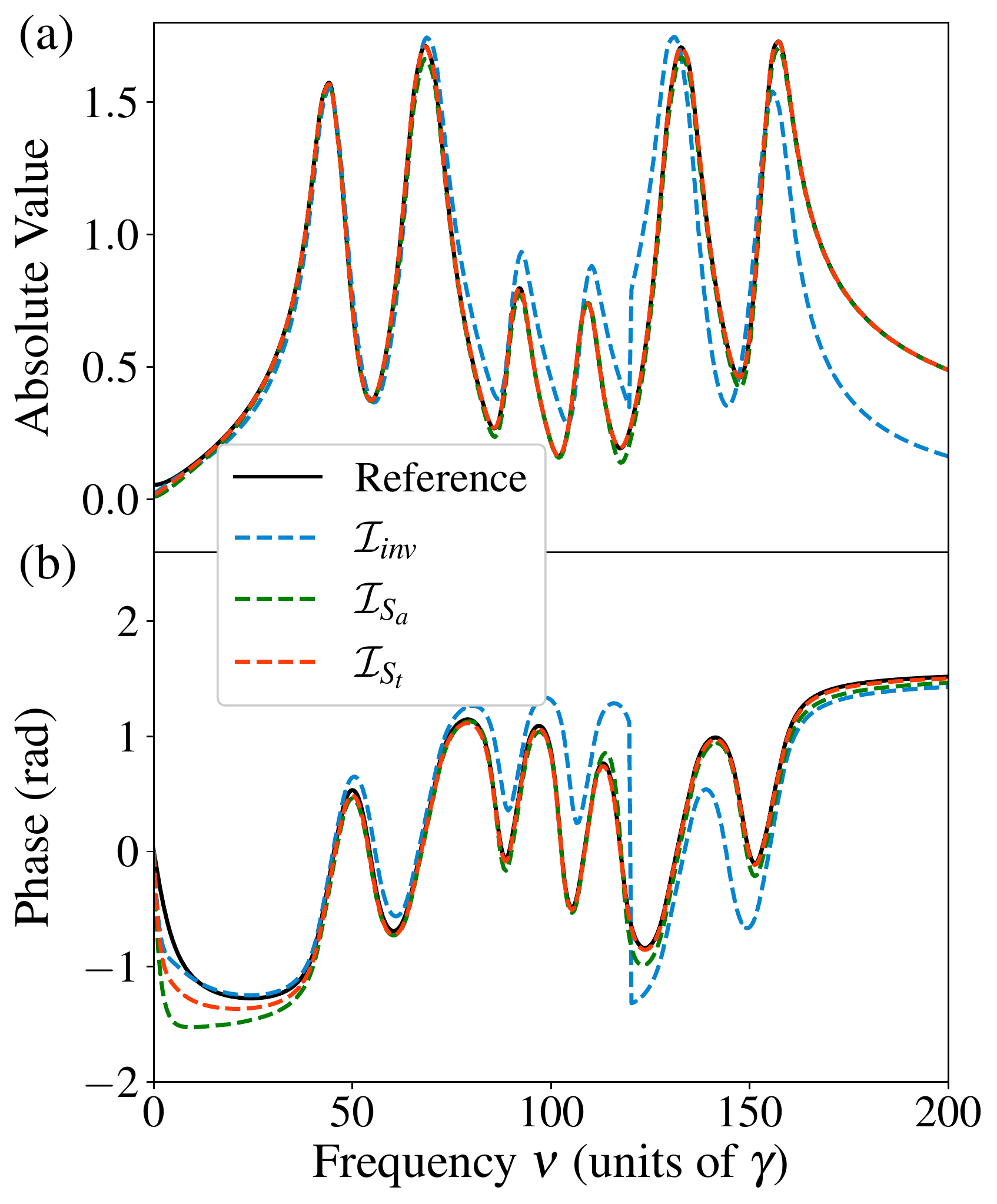}
\caption{\label{fig.CyclePhase} Reconstruction of the nuclear resonant target response from phase-combined \ffc spectra: (a) depicts the absolute value of a vertical cut through the \ffc spectra at $\Delta = 100\gamma$ (dashed lines). For comparison, the absolute value of the reference nuclear resonant response correlated with the analyzer response according to eq.(\ref{eq.OffResult}) is shown as solid black line. (b) shows the corresponding phase of the quantities shown in (a). The blue line corresponds to the background-corrected sum spectrum $\mathcal{I}_{inv}$ while the green and orange line correspond to the cases with the analyzer ($\mathcal{I}_{S_a}$) and the target ($\mathcal{I}_{S_t}$) placed first in the beam path, respectively.}
\end{figure}

We conclude our comparison of \ffc spectra obtained using phase-control with the off-resonant regime by reconstructing the complex-valued target response from sections along the $\nu$-axis as described in Section \ref{extract-nfs}.

Fig.~\ref{fig.CyclePhase} compares absolute value and phase of a vertical cut through the \ffc spectrum at $\Delta = 100 \gamma$ for the three phase combinations Eq.~(\ref{eq.combi-1})-(\ref{eq.combi-3}) with the reference spectrum obtained by cross-correlating nuclear resonant target and analyzer response explicitly (solid black line). It can be seen that the results based on the difference spectra $\mathcal{I}_{S_a}$ (orange line) and $\mathcal{I}_{S_t}$ (green line) agree very well with the theoretical reference. In particular for the case with  the target placed first in the beam path (orange line), the phase at lower frequencies $\nu$ is recovered even better than in the analysis based on the off-resonant regions of the \ffc spectrum in Fig.~\ref{fig.reconstruction}(b). Interestingly, the analysis based on the sum spectrum $\mathcal{I}_{inv}$ performs much worse, in particular towards higher frequencies. We attribute this to the background removal, which distorts the final result if resonant effects still contribute at the detuning values used to determine the background contributions (cf. discussion after Eq.~(\ref{eq.combi-3})).

Overall, we therefore conclude that additional control of the relative phase between the resonant and non-resonant radiation in Nuclear Resonant Scattering experiments can improve spectral analysis and target response reconstruction, as it allows one to selectively separate otherwise overlapping scattering channels across the entire detuning range. This improvement is manifest in a higher amount of data close to resonance that can be evaluated during the line fit analysis as well as a better recovery of the phase of the nuclear response in the lower Fourier frequency range compared to the off-resonant approach without phase control. Further, thickness effects impeding a reliable recovery of the spectral target structure via the diagonal lines can be suppressed by selecting appropriate scattering contributions thus allowing the study of thick targets that are not accessible to the off-resonant approach without phase control.

\section{\label{sec.summary}Discussion and Summary}

In this paper, we introduced frequency-frequency correlation (\ffc) spectra as a promising tool for spectral analysis and phase-retrieval of arbitrary M{\"o}ssbauer targets. These \ffc spectra are obtained  as Fourier transforms of experimentally accessible time- and frequency-resolved Nuclear Resonant Scattering spectra along the time-axis. Our approach is motivated by the observation that \ffc spectra exhibit a simple structure comprising horizontal and diagonal features which relate to different physical processes contributing to the scattered light, and which can conveniently be analyzed.

We showed that this approach translates interference between different scattering channels into frequency-frequency correlations revealing the spectral structure of the underlying scattering system. Specifically, the cross-correlation between nuclear resonant part of analyzer and target response appeared as a diagonal structure in these frequency-frequency correlation spectra that allowed access to target properties in two ways.

First, using linear fits to these diagonals, we were able to extract the resonance frequencies of the target, as well as spectral line features such as collective energy shifts and line broadenings, thereby  establishing an intuitive and straightforward analysis tool for the \ffc spectra.

Second, we found that sections through the diagonal structure provide access to amplitude and phase of the nuclear resonant part of the target response, cross-correlated with the analyzer response. In particular, they are not affected by the off-resonant electronic scattering in the target. This is of immediate interest for characterizing collective nuclear level schemes in x-ray cavities, which so far have been associated to  cavity reflection spectra~\cite{yoshida_quantum_2021}. However, these spectra also depend on the in- and out-coupling of the x-rays into the cavity and the interference of the nuclear response with the non-resonant empty-cavity scattering, and therefore the associated level schemes may not represent the actual nuclear dynamics inside the cavity~\cite{Diekmann2022}. Then it may be favorable to instead associate a level scheme to the nuclear dynamics inside the cavity alone, and the approach presented here allows one to experimentally access the corresponding nuclear response unperturbed by the electronic scattering contribution. In this sense, the analysis approach presented here has a qualitative advantage over the established late-time integration method. We note that time-domain off-resonant methods, as discussed in Sec.~\ref{sec:offres}, share this advantage, although so far they  have primarily been discussed in the forward scattering case~\cite{Callens2005,Goerttler2019}. This again highlights the advantage of time- and frequency-resolved measurements over time-resolved or frequency-resolved approaches, since they not only allow one to choose the late-time integration range in the data analysis after the actual experiment, but also provide the basis for both off-resonant analysis methods in the time- and energy-domain. In other words, a single dataset can be evaluated using three different methods, thereby also allowing for consistency checks.

To demonstrate the practical feasibility of both \ffc analysis approaches in relevant settings, we employed them to determine magnetic hyperfine splittings of $\alpha$-iron in forward scattering geometry as well as the collective Lamb shift of thin-film x-ray cavities as function of incidence angle $\theta$ from simulated data. In all cases, good agreement was achieved with corresponding reference calculations. Further, the superradiant enhancement of the nuclear decay in the cavity could successfully be retrieved using an appropriate fit model. We also considered experimental resolution limitations along the Fourier frequency $\nu$ in \ffc spectra, which may arise due to measurement constraints in the time domain, e.g., due to the x-ray pulse repetition rate determined by the radiation source. This can be mitigated by employing our approach based on the  relative detuning $\Delta$ between analyzer and target rather than on $\nu$, since the former has an experimentally controllable resolution independent of the x-ray pulse structure. An example analysis of the hyperfine splitting of $\alpha$-iron including time gating effects indeed showed that this way, results with reasonable accuracy could be obtained.

In a second part, we considered the possibility of further extending the \ffc approach based on a control of the relative phase between the resonant and off-resonant analyzer response. In the presence of such phase-control, the overall response becomes dependent on the ordering of target and analyzer. We showed that sums or differences of \ffc spectra recorded with suitable phase shifts or analyzer-target orderings allow one to disentangle different scattering pathways across the entire \ffc spectrum. In contrast, most previous approaches separated the pathways by restricting the analysis to particular parameter regions such as the large target-analyzer detuning case. This way, spectral backgrounds can be removed, or scattering paths be excluded which prevent an accurate \ffc analysis, especially close to the resonances. In the present work, we employed this technique to isolate the interference contribution between the individual resonant target and analyzer scatterings. We could thereby improve the recovery of phase and spectral target structure as compared to the off-resonant case without phase control and reliably retrieve the target's resonance structure even for thick targets where the off-resonant limit is difficult to reach. In addition, we expect that the approach will also facilitate the study of transition-specific dynamics and resonant couplings~\cite{PhysRevA.65.023804,PhysRevA.71.023804,Heeg2021} by suppressing non-resonant contributions to time- and frequency-resolved spectra.

\appendix

\section{\label{app:late}Spectroscopy via late-time integration}

The late-time integration spectroscopy method also employs frequency-tunable reference absorbers and has proven to be particularly successful in quantum optical studies involving thin-film cavities \cite{Roehlsberger2010,Roehlsberger2012,Heeg2013SGC,Heeg2015,Heeg2015SL,haber_collective_2016,haber_rabi_2017}. To gain insight into its main assumptions, we briefly recall the derivation of the late-time method given in Ref.~\onlinecite{Roehlsberger2010}. The experimentally accessible time- and frequency-resolved intensity $I(t,\Delta)$ as given in Eq.~(\ref{eq.tf-int}) is integrated for late photon arrival times $[t_1,t_2]$ yielding the frequency-resolved observable
\begin{align}
\mathcal{I}_{\ltm}(\Delta) = E^2_0 \lvert \alpha_a\rvert^2 \int^{t_2}_{t_1} dt \lvert T_t(t) +(T_t \ast S_a)(t)\rvert^2
\end{align}
where the detuning dependence enters via the Doppler-shifted analyzer response $T_a(t) = \alpha_a\left[\delta(t) +S_a(t,\Delta) \right]$.
If the nuclear resonant part of the analyzer response $S_a$ is only nonzero in the vicinity of $\omega = \omega_a + \Delta$ and the target response varies slowly over this frequency range, their convolution can be written as
\begin{align}
(T_t \ast S_a)(t) &= \mathcal{F}^{-1}\left[\hat{T}_t(\omega)\hat{S}_a(\omega,\Delta)\right] \nonumber \\
&\approx \hat{T}_t(\omega_a+\Delta) S_a(t,\Delta) \,.
\end{align}
This approximation can be extended straightforwardly to include first order derivatives of the phase of $\hat{T}_t$ (cf. \cite{Herkommer2020, HeegPhD}) but in general requires that the target response is approximately constant across the spectral range of the analyzer. Assuming further that at late times $t_1\gg0$ the target response $T_t(t)$ has already decayed, one can further approximate
\begin{align}
\mathcal{I}_{\ltm}(\Delta) \approx E^2_0 \lvert \alpha_a\rvert^2 \mathcal{C} \lvert \hat{T}_t(\omega_a+\Delta)\rvert^2\,, \label{eq.LateTime}
\end{align}
where $\mathcal{C} = \int^{t_2}_{t_1}dt \lvert S_a(t)\rvert^2$. Thus, for sufficiently late integration times  the integrated intensity $\mathcal{I}_{\ltm}$ becomes proportional to the absolute squared of the frequency-domain target response $\lvert \hat{T}_t(\omega_a+\Delta)\rvert^2$.

The two key approximations performed above require target spectra which are spectrally broad and smooth on scales of the spectral analyzer width. Therefore, the method is particularly useful for thin-film cavities probed in reflection which feature spectrally broad resonances.

\begin{figure}[t]
\includegraphics[width = 0.5 \textwidth]{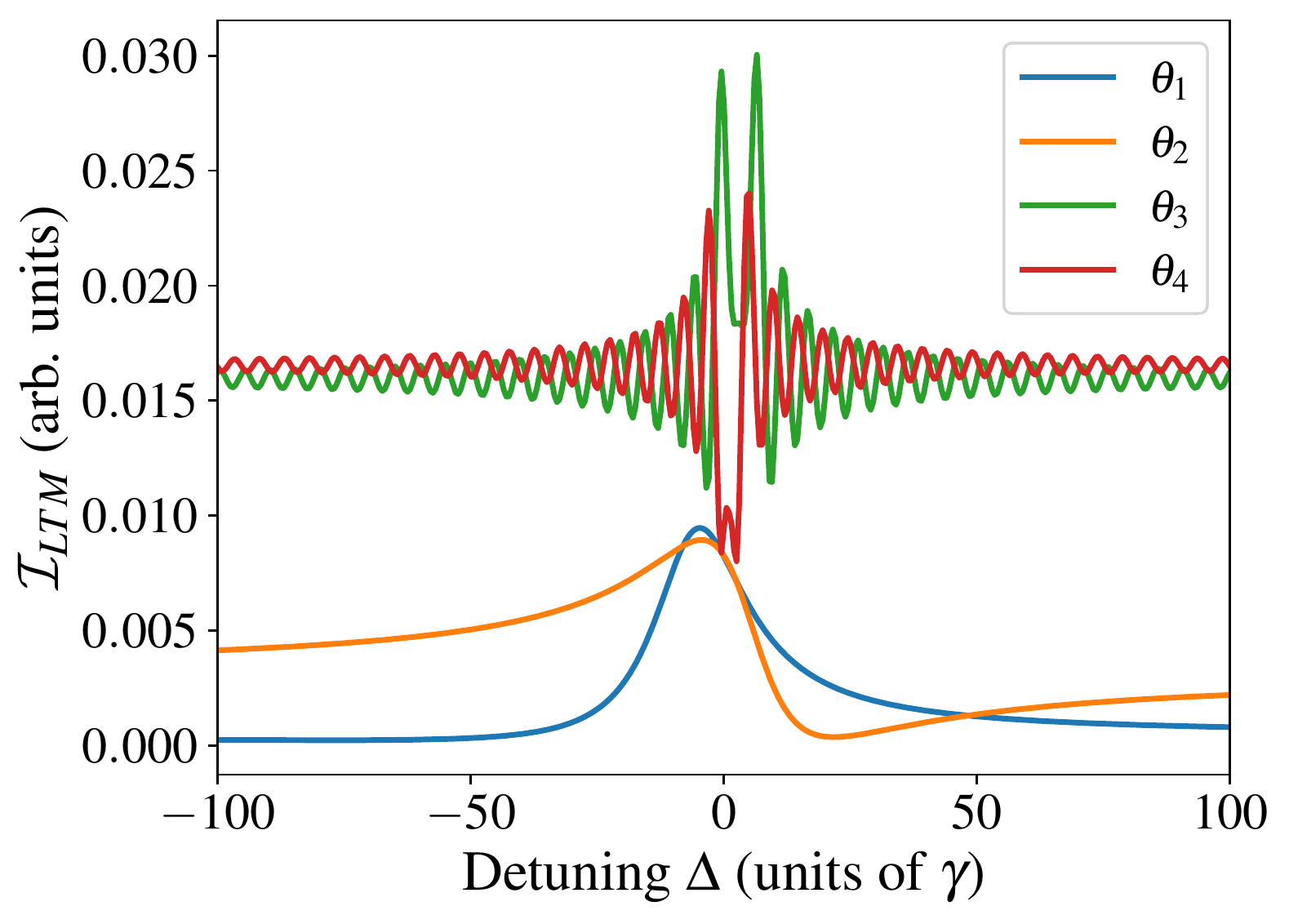}
\caption{\small{Recovery of cavity-spectra using late-time integration. The Figure shows spectra obtained by integrating the time- and frequency-resolved spectra over the time interval between $t_1 = 180$ns and $t_2 = 1600$ns. The four curves correspond to incidence angles $\theta_1 = 2.75$mrad, $\theta_2 = 2.8$mrad, $\theta_3 = 3.0$mrad and $\theta_4 =3.3$mrad. The cavity is the same as the one considered in Sec.~\ref{subsec.Cav}, and the intensity scale is as in Fig. \ref{fig.CavityPlots}(a). At incidence angles $\theta_1$ and $\theta_2$, the nuclear resonant response is spectrally broad, and the spectra are reproduced well by the late-time integration, though with reduced visibility. In contrast, the late-time spectra at incidence angles $\theta_3, \theta_4$ suffer from artefacts of the time integration and do not represent the refrence spectra in Fig. \ref{fig.CavityPlots}  well. \label{fig.LateTime}}}
\end{figure}

The strengths of the late-time method lies in its simple measurement and analysis approach. However, in comparison to the \ffc methods discussed here, the late-time integration has several drawbacks. First, it does not allow one to retrieve the complex-valued target response. Second, it relies only on data recorded at late times, such that most of the signal photons cannot be used for the spectrum recovery. Finally, the recovery sensitively depends on the choice of $t_1$ and $t_2$, and the obtained spectra may be perturbed by time gating effects.
Some of these challenges are illustrated in Fig.~\ref{fig.LateTime}, which shows late-time spectra of a thin-film cavity with layer structure as given in Sec. \ref{subsec.Cav} for the incidence angles considered in Fig.\ref{fig.CavityPlots} with integration boundaries of $t_1 = 180$ns, $t_2 = 1600$ns.

At incidence angles $\theta_1$ and $\theta_2$, the target spectra are reproduced well. However, the neglect of the early photons causes a reduction of the usable spectral intensity by about three orders of magnitude, as can be seen in comparison to the reference spectra Fig.\ref{fig.CavityPlots}, which are drawn using the same scale. Further, the late-time spectra for incidence angles $\theta_3$ and $\theta_4$ are completely distorted, even though only the incidence angle was changed. The reason for this is that at these incidence angles, the cavity resonance line-width is small (cf. Fig.~\ref{fig.CavParams}(b)), such that the approximations underlying the late-time integration method break down. Note that time- and frequency-resolved data acquisition can partially compensate for this downside since it allows for  \textit{a posteriori} selection of the integration time window.

\section{Collective parameters in cavity reflection spectra \label{sec.ColParams}}

In order to retrieve collective nuclear parameters such as the collective Lamb shift $\Delta_{CLS}$ and superradiance $\gamma_s$ from the diagonal structure appearing in \ffc spectra, a model function is required to fit the section along the $\nu$-axis through the diagonals shown in Fig.~\ref{fig.CavityPlots}. 

To obtain such a fit function, we employ a quantum optical model~\cite{Heeg2013Model} of M{\"o}ssbauer nuclei embedded in thin-film cavities, which allows one to derive analytical expressions for these parameters. For this, we calculate the time-domain response for grazing-incidence reflection from a thin-film cavity with a single layer of M{\"o}ssbauer nuclei embedded in its center (cf. Fig.\ref{fig.Setup}) by Fourier transforming the cavity's frequency-domain response~\cite{HeegPhD}

\begin{align}
T_t(\omega) = \alpha_t - i(\alpha_t+1)\frac{\left(\frac{\gamma_s}{2}+i\Delta_{CLS}\right) }{\omega-(\omega_t+\Delta_{CLS})+i\Gamma_c}
\end{align}
where the empty-cavity reflection coefficient is given by 

\begin{align}
 \alpha_t = \frac{2\kappa_R}{\kappa+i\Delta_c}-1\,.
\end{align}

Here, $\kappa$ is the total loss rate of the cavity, $\Delta_c$ the detuning between the cavity mode and an external driving mode, and $\kappa_R$ is the in- and outcoupling rate of the cavity~\cite{Heeg2013Model}. $\Gamma_c = (\gamma+\gamma_s)/2$ denotes the total decay rate of the nuclei in the cavity with $\gamma_s$ the superradiant line-width~\cite{dicke_coherence_1954,gross_superradiance_1982,garraway_dicke_2011,Roehlsberger2010,chumakov_superradiance_2018,HannonTrammell1999,lynch_time_1960,kagan_excitation_1979,guerin_light_2017}, and $\Delta_{CLS}$ a level shift experienced by the nuclear ensemble known as collective Lamb shift~\cite{Roehlsberger2010,longo2016tailoring,shv,friedberg_frequency_1973,scully_collective_2009,ruostekoski_emergence_2016,keaveney_cooperative_2012,peyrot_collective_2018,roof_observation_2016,bromley_collective_2016}. Though in general frequency-dependent through the cavity-detuning $\Delta_c$, we regard $\Delta_{CLS}$, $\gamma_s$ and $\alpha_t$ as constants as function of frequency since the cavity resonances are typically orders of magnitude broader than nuclear resonances \cite{Lentrodt2020} and thus the empty-cavity part becomes essentially flat in frequency-space. In this case, a Fourier transform of the frequency response yields:

\begin{align}
T_t(t) =& \alpha_t \delta(t) - (\alpha_t+1)\left(\frac{\gamma_s}{2}+i\Delta_{CLS}\right)  \nonumber \\
&\qquad \times e^{-i(\omega_t+\Delta_{CLS})t}e^{-\Gamma_ct}\Theta(t) \,,\label{app:cavity}
\end{align}

Employing the early-time approximation Eq.~(\ref{app:eq2}) for the analyzer response, 
the \ffc spectrum in the large analyzer-target detuning limit Eq.~(\ref{eq.OffResult}) for a cavity target given by  Eq.~(\ref{app:cavity}) can be calculated in the time domain as
\begin{align}
\FourierInt{} =& E^2_0 \int^{\infty}_{0}dt e^{i\nu t}e^{-S_{\Gamma}t} \left[ce^{-ixt}+c^*e^{ixt}\right] \nonumber\\[1ex]
= & iE^2_0 \left[\frac{c}{\nu-x+iS_{\Gamma}}+\frac{c^*}{\nu+x+iS_{\Gamma}}\right] \nonumber \\[1ex]
= & 4E^2_0\left[\frac{\textrm{Re}(c)(i\nu-S_{\Gamma})-\textrm{Im}(c)x}{(\nu+iS_{\Gamma})^2-x^2}\right]  . \label{eq.CavSpectrum}
\end{align}
Here, we have introduced the parameters 
\begin{subequations}
\label{app:coeff}
\begin{align}
 S_{\Gamma} &= \Gamma_c+\frac{\gamma_a+b_a}{2} \,,\\
 x &= \Delta-(\Delta_{CLS}+\omega_t-\omega_a)\,,\\
 c &=  \alpha_t(\alpha^*_t+1)\left(\frac{\gamma_s}{2}-i\Delta_{CLS}\right)b_a\,.
\end{align}
\end{subequations}
Note that evaluating the absolute value squared of the complex-valued \ffc spectrum Eq.~(\ref{eq.CavSpectrum}) yields two Lorentzians in the limit of large detunings, 
\begin{align}
\lvert \FourierInt{}\rvert^2 = & 16 E^4_0 \left(\frac{\lvert c\rvert^2}{(\nu-x)^2+S^2_{\Gamma}}+\frac{\lvert c\rvert^2}{(\nu+x)^2+S^2_{\Gamma}}\right. \nonumber \\
& \left. + 2 \textrm{Re}\left[\frac{c^2}{(\nu-x+iS_{\Gamma})(\nu+x-iS_{\Gamma})}\right]\right) \nonumber \\
\xrightarrow[x \gg S_{\Gamma}]{} & 16 E^4_0 \lvert c\rvert^2 \left(\frac{1}{(\nu-x)^2+S^2_{\Gamma}}+\frac{1}{(\nu+x)^2+S^2_{\Gamma}}\right) . \nonumber
\end{align}
This is consistent with our observation of a Lorentzian line shape in diagonal cavity spectra in Sec.~\ref{subsec.Cav}, and further supports our key result that the diagonal cavity spectrum indeed yields the nuclear resonant part $S_t$ of the spectrum only. This is most easily seen by the fact that spectra affected by the interference of the nuclear response with the electronic background response in general have Fano line shapes, see Fig.\ref{fig.CavityPlots}.  As in the case of nuclear forward scattering, the response is shifted in the \ffc spectrum by the analyzer-target detuning via $x$, and broadened by the analyzer response width via $S_{\Gamma}$.

In order to retrieve collective nuclear parameters from vertical cuts through the diagonal structure, we define a more general fit function based on  the absolute value of Eq.~(\ref{eq.CavSpectrum}) as
\begin{align}
D(\nu) = \lvert \FourierInt{}\rvert = A \sqrt{\frac{\nu^2 +(px)^2}{(\nu^2-x^2+S^2_{\Gamma})^2+4x^2S^2_{\Gamma}}} \label{eq.DiagonalFit}
\end{align}
where we assumed that $\frac{S_{\Gamma}}{x}\ll 1$,  which is consistent with the off-resonant approximation used to derive $\FourierInt{}$. The fit parameters related to Eqs.~(\ref{app:coeff}) are $A = 4 E^2_0 \textrm{Re}(c)$ and $p = \textrm{Im}(c)/\textrm{Re}(c)$.\\
The position of the maximum of each vertical section is given by $x$ and the lines formed by these maxima can be fitted similarly to the procedure described in Section \ref{sec.expar} to obtain $\Delta_{CLS}$ as crossing point with the $\nu$-axis. The total decay rate $\Gamma_c$ is determined by averaging over the $S_{\Gamma}$ retrieved from each vertical section and subsequent subtraction of the effective analyzer width $\frac{\gamma_a+b_a}{2} = 1.67 \gamma_a$ for a $1\mu$m thick stainless steel analyzer containing $55\%$ of $^{57}\textrm{Fe}$. \\
Since cavity spectra can be considerably broadened by superradiance, we use two optimization strategies to reduce resonant effects: First, we optimize the $\Delta$ fit range towards the lowest line fit error as described in Section ~\ref{sec.expar}, however, by evaluating the first 40 fits. This suppresses resonant effects affecting the shape of the diagonal line at small detunings while at the same time taking into account as many data points as possible. The second optimization is related to low-frequency deviations from the Lorentzian line shape also caused by resonant effects: To minimize its influence on the fit outcome, a series of model fits given by eq.(\ref{eq.DiagonalFit}) with adapted $\nu$ fit range is performed on each vertical cut taking into account a different number of data points below the peak maximum. Specifically, we first include a frequency range starting right below the expected peak maximum only and evaluate eight fits $j \in \{0, 1, \dots, 7\}$ in which $10j$ datapoints are added at the lower end of the vertical cut. Again, the fit with the least fit error is selected. It was found that this approach improves the fit result especially in the vicinity of the superradiance maximum where resonant effects are expected to be strongest.

\section{Line shapes in Nuclear Forward Scattering \label{sec.NFSLines}}

In this Appendix, we briefly discuss the response functions of single-line targets in nuclear forward scattering in the time domain, and employ them to evaluate their line shape in the \ffc spectra.

For a single-line target, the scattering part of the generic response function in the time domain Eq.~(\ref{eq:gen-time}) evaluates to~\cite{lynch_time_1960,kagan_excitation_1979,Smirnov1999,Shvydko1998, RoehlsbergerBook,HannonTrammell1999}
\begin{align}
S_t(t)=-\Theta(t) \: \sqrt{\frac{b_t}{t}} \,  J_1(2\sqrt{b_t t}) e^{-\frac{\gamma_t}{2}t}\, e^{-i\omega_t t}\,,\label{app:eq1}
\end{align}
where $\Theta(t)$ is the Heaviside step function, $\gamma_t$ the decay rate of the individual nuclei, and $\omega_t$ the transition frequency.  The thickness parameters $b_i$ for target ($i=t$) and analyzer ($i=a$) are given by
\begin{align}
b_i = \frac{\pi\rho_i f_{LM} \gamma_i}{k^2_i(1+\alpha)}d_i \label{eq.TheoThick}
\end{align}
where $\rho_i$ denotes the number density of resonant nuclei, $\gamma_i$ its line-width, $k_i$ the wave number of the resonant radiation, $d_i$ the target's thickness, $f_{LM}$ the Lamb-M{\"o}ssbauer factor and $\alpha$ the internal conversion coefficient (for more details, see e.g. \cite{HannonTrammell1999, RoehlsbergerBook}). For $\alpha$-iron the internal conversion coefficient is given by $\alpha = 8.56$ \cite{RoehlsbergerBook} and the Lamb-M{\"o}ssbauer factor at ambient conditions is of the order $f_{LM} \approx 0.8$ \cite{Sturhahn2004}.\\

In the early-time approximation $b_a t\ll 1$, the corresponding response for a thin  single-line analyzer evaluates to
\begin{align}
S_a(t,\Delta) = -\Theta(t)\: b_a \, e^{-\frac{\gamma_a+b_a}{2}t}\: e^{-i(\omega_a+\Delta)t} \, . \label{app:eq2}
\end{align}
This early-time approximation of the analyzer response is based on 
\begin{align}
\sqrt{\frac{b_a}{t}}J_1(2\sqrt{b_at}) \approx b_ae^{-\frac{b_a}{2}t} \label{app.Early}
\end{align}
which is valid for $b_a t \ll 1$ up to first order.

The two response functions Eq.~(\ref{app:eq1}) and (\ref{app:eq2}) can be used to evaluate the expression for the \ffc spectrum in Eq.~(\ref{eq.OffResult}) in the time domain. Then, the Fourier transform of the positive branch (also c.f. Eq.~(\ref{eq.CrossCorr})) is given by
\begin{align}
f(\nu) =& \IZero\int^{\infty}_{-\infty}\,dt \: e^{i\nu t}\: S^*_t(t)\,S_a(t)\nonumber \\[1ex]
=& \IZero \: b_a\: \int^{\infty}_{0}\:dt\: \sqrt{\frac{b_t}{t}}e^{i(\nu-x)t}e^{-S_{\Gamma}t}J_1(2\sqrt{b_t t}) \nonumber\\[1ex]
=& A \left[\exp\left(\frac{-ib_t}{\nu-x+iS_{\Gamma}}\right)-1\right] \label{app:eq3}
\end{align}
with
\begin{align}
x& =\Delta-\omega_t+\omega_a \,,\\
S_{\Gamma} &= \frac{\gamma_t+\gamma_a+b_a}{2}\,,\\
	A &= b_a\IZero\,.
\end{align}
Since the line shapes obtained from the \ffc spectrum coincide with the target response functions as shown in the main text, Eq.~(\ref{app:eq3}) also corresponds to the well-known frequency response function in nuclear forward scattering, as expected. However, as compared to the standard expressions~\cite{Smirnov1999}, Eq.~(\ref{app:eq3}) contains a spectral broadening entering in $S_{\Gamma}$ caused by the convolution with the analyzer, and an additional frequency shift due to the interference with the analyzer radiation leading to the diagonal structure of the \ffc spectrum.

In the main text, we employ Eq.~(\ref{app:eq3})  as the fit model in Eq.~(\ref{eq.ThickModel}) to extract spectral parameters from the phase-combined and off-resonant spectra.

\section{Phase-dependence of fields and intensities upon reversing sample order \label{sec.TargInv}}

In this appendix, we derive the x-ray field behind a combination of two nuclear absorbers, both of which acquire sudden relative phase shifts immediately after arrival of the x-ray pulse, either by near-instantaneous displacements or by rapid changes of the magnetic field at the nuclear positions (for details, see Sec.~\ref{sec.PhaseCycle}). In particular, we focus on the effect of the ordering of the two absorbers.

We start by deriving the field behind a single absorber. In order to evaluate the response of the absorber, it is convenient to transform into its rest frame, which refers to the frame where the nuclear resonances are time-independent. Denoting the lab frame incident field $E^L_{in}(t)$, the corresponding incident field in the rest frame is
\begin{align}
E^R_{in}(t) = e^{-i\phi(t)}E^{L}_{in}(t)\,,
\end{align}
where $\phi(t)$ is the time-dependent phase shift. The outgoing field in the rest frame is
\begin{align}
E^{R}_{out}(t) &= \left(T \ast E^{R}_{in}\right)(t) \nonumber\\[1ex]
&= \int^{\infty}_{-\infty}dt' \: T(t-t')  \: e^{-i\phi(t')} \: E^{L}_{in}(t') \,.
\end{align}
Finally, transforming back into the lab frame yields
\begin{align}
E^{L}_{out}(t) =& e^{i\phi(t)}E^{R}_{out}(t)\nonumber \\[1ex]
=& \int^{\infty}_{-\infty}dt' \: e^{i[\phi(t)-\phi(t')]}\: T(t-t')\: E^{L}_{in}(t')\,.
\end{align}
Next, we calculate the field behind two absorbers, the nuclear resonances of both change with time. We denote the response function of the upstream [downstream] absorber as $T_1$ [$T_2$], with phase shifts given by $\phi_1$ [$\phi_2$]. We obtain
\begin{align}
E_{out}(t) =&  \int^{\infty}_{-\infty}dt' \int^{\infty}_{-\infty}dt''\: e^{i[\phi_2(t)-\phi_2(t')]}\: T_2(t-t')\times\nonumber\\[1ex]
& \: e^{i[\phi_1(t')-\phi_1(t'')]}\: T_1(t'-t'')\: E^{L}_{in}(t'')\,.
\end{align}
Assuming a near-instantaneous incident field $E^{L}_{in}(t'') = E_0\delta(t)$, we find
\begin{align}
E_{out}(t) =& E_0 \int^{\infty}_{-\infty}dt' \: e^{i[\phi_2(t)-\phi_2(t')]}\: T_2(t-t')\times\nonumber\\[1ex]
& \qquad e^{i[\phi_1(t')-\phi_1(0)]}\: T_1(t')\nonumber\\[1ex]
=& E_0\left[ \delta(t) +  e^{i[\phi_1(t)-\phi_1(0)]}\: S_1(t) + \right.\nonumber\\[1ex]
& +  e^{i[\phi_2(t)-\phi_2(0)]}\: S_2(t) \nonumber\\[1ex]
 &+ \int^{t}_{0}dt' e^{i[\phi_2(t)-\phi_2(t')]}\: e^{i[\phi_1(t')-\phi_1(0)]} \times\: \nonumber\\[1ex]
 &\qquad  S_2(t-t') S_1(t') \Bigr]  \label{eq.TwoJump}
\end{align}
Here, we have used $T(t)=\delta(t)+S(t)$, neglected for simplicity the non-resonant electronic absorption, and the integral ranges are constrained since $S(t)\propto \Theta(t)$, where $\Theta(t)$ is the Heaviside unit step function.
As expected, the individual resonant responses of the two absorbers are modified by their respective phase shift, as can be seen from the second and third term in in Eq.~(\ref{app-d-1}). However, the last term, which corresponds to the radiative coupling, is only affected by the motion of the upstream absorber in case of near-instantaneous phase-jumps. This is due to the fact that the effect of the step-like shift of the second absorber can be neglected since  $\exp(i[\phi_2(t)-\phi_2(t')])$ in the radiative coupling term contributes only in the negligible interval $t'=0$. Intuitively, only relative phase shifts between in- and outgoing radiation may modify the outgoing field, and all x-rays scattered by the upstream absorber reaching the second absorber after the incident pulse at $t=0$ are not affected by its change in phase since it is already finished. For analogous reasons, the contribution of the step-like phase $\phi_1$ can be moved out of the integral, since $\phi_1(t') = \phi_1(t'')$ for any $t',t''>0$. With this, Eq.(\ref{eq.TwoJump}) can be written as

\begin{align}
E_{out}(t) =& E_0\left[ \delta(t) +  e^{i[\phi_1(t)-\phi_1(0)]}\: S_1(t) + \right.\nonumber\\[1ex]
& +  e^{i[\phi_2(t)-\phi_2(0)]}\: S_2(t) \nonumber\\[1ex]
 &+ e^{i[\phi_1(t)-\phi_1(0)]}  S_{1,2}(t) \Bigr]\,, \label{app-d-1}
\end{align}

where we have defined the radiative coupling contribution
\begin{align}
S_{1,2}(t) = \int^{\infty}_{-\infty}dt' S_2(t-t')S_1(t')\,.
\end{align}

Using Eq.~(\ref{app-d-1}), adding the absorption and dispersion prefactors $\alpha$, assuming $\phi_1(0)=0=\phi_2(0)$ without loss of generality, and specifying the ordering of analyzer and target, immediately leads to Eqs.~(\ref{eq.AnaFirst})-(\ref{eq.TargFirst}) of the main text.

Finally, the intensity after the two absorbers follows from Eq.~(\ref{app-d-1}) as
\begin{align}
I(t) =& I_b + 2I_0\textrm{Re}\left[e^{i(\phi_2-\phi_1)}S^*_1S_2\right. \nonumber \\[1ex]
&\left.+S^*_1S_{1,2}+e^{i(\phi_2-\phi_1)}S_2S^*_{1,2}\right]\,, \label{eq.phaseint-final}
\end{align}
where we have again assumed $\phi_1(0)=0=\phi_2(0)$  and defined $\phi_i \equiv \phi_i(t>0)$. From this result, two important conclusions can be drawn. First, phase control of both absorbers does not add an additional degree of freedom beyond the phase control of only one absorber, since only the difference of the two phases enters the expression for the experimentally accessible intensity. Hence, we restrict our discussion to the phase control of the analyzer, since a target phase control may not always be experimentally feasible, i.e., for thin-film cavities.
Second, if only one of the absorbers (analyzer) is moved, then the ordering of the two absorbers is crucial.
The corresponding intensities for either the first or the second absorber being moved are given in Eqs.~(\ref{eq.PhaseImprint}) of the main text.

\bibliography{Diagonal}

\end{document}